# Primary Period–Luminosity-Relation Calibrators in the Milky Way: Cepheids and RR Lyrae

## Physical basis, Calibration, and Applications

### M. A. T. Groenewegen

Koninlijke Sterrenwacht van België, Ringlaan 3, 1180 Brussels, Belgium
martin.groenewegen@oma.be

**Abstract.** In this invited review I discuss the calibration and applications of the period–luminosity relation of classical Cepheid and RR Lyrae stars. After a brief introduction, starting with results from *Hipparcos* and discussing some post-*Hipparcos* era developments, I focus on recent results using *Gaia* Data Release 3 data. I present an overview of the most recent period–luminosity relations, a discussion and some new results on Cepheids in open clusters. I also discuss the effect of reddening and that the use of Wesenheit indices is actually an oversimplification to dealing with the problem of reddening.

## 1. Introduction

Period–luminosity (PL) relations are key tools to determine distances to classical variables like RR Lyrae (RRL) and classical Cepheids (but also to Type II Cepheids (T2Cs), Mira and $\delta$ Scuti variables), and therefore, tracers of Galactic structure. The calibration of these PL relations is an important step, and even a crucial one in, arguably, the most important application of PL relations, i.e., anchoring the local distance scale, in particular for classical Cepheids (DCEPs).

In this review I invert the order of my talk in Budapest and start by discussing the calibration of PL relations, focussing on recent results obtained with *Gaia* Data Release 3 (DR3) data (Gaia Collaboration et al. 2021, 2023b), and I then discuss applications to Galactic structure. A similar review was given at IAU Symposium 330 (Groenewegen 2018a) when *Gaia* DR1 had just become available (Gaia Collaboration et al. 2016b,a).

### 1.1. *The Hipparcos and pre-Gaia era*

Let us turn back the clock a quarter of a century. *Hipparcos* provided parallax and proper motion (PM) data for a predefined sample of about 110,000 objects (ESA 1997). Feast and Catchpole (1997) derived the PL relation for DCEPs in the *V* band†. *Hipparcos* data were available for 220 DCEPs, although most weight was in a subset of 26 stars, which itself was dominated by one Cepheid, $\alpha$ UMi (Polaris). Although revolutionary at the time, the parallax data were not good enough to determine the slope of the PL relation (it was fixed to the value derived for the LMC), let alone to explore any metallicity dependence (which was not considered).

They were also the first to work in 'parallax space', avoiding any issues with taking the logarithm of a distance in case of negative parallaxes. The (photometric) parallax is defined as

$$\pi = 10^{0.2(m_0 - F - 5)},\tag{1}$$

† $M_V = -2.81 \log P + (-1.43 \pm 0.10)$

where $\pi$ is the parallax, $m_0$ the dereddened magnitude, and $F$ any (non-)linear function of period, metallicity, or any other quantity. This equation can be solved using maximum-likelihood techniques or $\chi^2$ minimisation for the parameters of the function $F$, taking into account the errors in the parallax and the observables.

Shortly after Feast and Catchpole (1997), Madore and Freedman (1998) used $BVIJHK$ data of 8–19 Cepheids (depending on the combination of available magnitudes) to determine reddening values and the distance to the LMC†. Ten years later, van Leeuwen et al. (2007) used the same sample of 220 DCEPs to derive a PL relation in the $K$ band‡.

For RR Lyrae stars, even fewer (accurate) *Hipparcos* parallaxes were available and Gratton (1998) used 3 RR Lyrae, 9 red horizontal branch (HB), and 10 blue HB stars to derive a relation between $V$-band magnitude and metallicity§.

A step forward in parallax accuracy came about when it was realised that the *Hubble Space Telescope* (*HST*) could be used to determine parallaxes. A paper on the namesake RR Lyrae was published by Benedict et al. (2002), followed by papers on nine DCEPs (Benedict et al. 2007), five RRL, and two T2Cs (Benedict et al. 2011), and the Cepheid $\eta$ Aql (Benedict et al. 2022). Over one hundred objects have had their parallaxes determined using the *HST*/Fine Guidance Sensor (FGS); see the review by Benedict et al. (2017). Then, Riess et al. (2014) developed a spatial scanning technique using the Wide-Field Camera 3 (WFC3) on board the *HST* to derive the parallax of SY Aur, and of seven other DCEPs (Riess et al. 2018a).

**Table 1.** Cepheid parallaxes (in mas) from *Hipparcos* to *Gaia*

| Name | $\pi \pm \sigma_\pi$ (rev. *Hipparcos*) | $\pi \pm \sigma_\pi$ (*HST*) | $\pi \pm \sigma_\pi$ (GDR3) | GoF |
|---|---|---|---|---|
| $\eta$ Aql | $2.36 \pm 1.04$ | $3.71 \pm 0.07$ | $3.67 \pm 0.03$ | 29 |
| $\beta$ Dor | $3.24 \pm 0.36$ | $3.14 \pm 0.16$ | $2.93 \pm 0.14$ | 47 |
| $\delta$ Cep | $3.77 \pm 0.17$ | $3.66 \pm 0.15$ | $3.56 \pm 0.15$ | 31 |
| FF Aql | $2.11 \pm 0.33$ | $2.81 \pm 0.18$ | $1.91 \pm 0.07$ | 1.5 |
| $l$ Car | $2.09 \pm 0.29$ | $2.01 \pm 0.20$ | $1.98 \pm 0.11$ | 24 |
| RT Aur | $-1.10 \pm 1.41$ | $2.40 \pm 0.19$ | $1.81 \pm 0.12$ | 66 |
| T Vul | $2.71 \pm 0.43$ | $1.90 \pm 0.23$ | $1.69 \pm 0.06$ | 5.3 |
| Y Sgr | $2.64 \pm 0.45$ | $2.13 \pm 0.29$ | $1.97 \pm 0.06$ | 13 |
| X Sgr | $3.31 \pm 0.26$ | $3.00 \pm 0.18$ | $2.81 \pm 0.14$ | 5.1 |
| $\zeta$ Gem | $2.37 \pm 0.30$ | $2.78 \pm 0.18$ | $3.07 \pm 0.22$ | 22 |
| W Sgr | $3.75 \pm 1.12$ | $2.28 \pm 0.20$ | $2.37 \pm 0.18$ | 36 |
| | | | | |
| SY Aur | $-1.84 \pm 1.72$ | $0.428 \pm 0.054$ | $0.427 \pm 0.020$ | 2.5 |
| SS CMa | $0.40 \pm 1.78$ | $0.389 \pm 0.029$ | $0.287 \pm 0.013$ | 4.0 |
| XY Car | $-1.02 \pm 0.88$ | $0.438 \pm 0.048$ | $0.378 \pm 0.014$ | 2.1 |
| VX Per | $0.87 \pm 1.52$ | $0.420 \pm 0.076$ | $0.364 \pm 0.017$ | 3.5 |
| VY Car | $0.36 \pm 1.42$ | $0.586 \pm 0.045$ | $0.554 \pm 0.017$ | $-2.4$ |
| WZ Sgr | $3.50 \pm 1.22$ | $0.512 \pm 0.039$ | $0.574 \pm 0.028$ | $-1.1$ |
| S Vul | | $0.322 \pm 0.040$ | $0.205 \pm 0.020$ | 1.1 |
| X Pup | $1.97 \pm 1.26$ | $0.277 \pm 0.048$ | $0.376 \pm 0.020$ | 1.2 |

Notes. Col. 2: Revised *Hipparcos* parallaxes (van Leeuwen 2007, 2008); Col. 3: *HST* parallaxes – Benedict et al. (2022) for $\eta$ Aql, Benedict et al. (2007) for $\beta$ Dor/W Sgr, Riess et al. (2014) for SY Aur, and Riess et al. (2018a) for SS CMa/X Pup; Col. 4: GDR3 parallaxes, with the goodness-of-fit (GoF) parameter in Col. 5.

† $E(B-V)$ ranging from $0.14 \pm 0.08$ mag to $0.17 \pm 0.07$ mag and distance moduli ranging from $18.44 \pm 0.35$ mag to $18.57 \pm 0.11$ mag.

‡ $M_K = -3.258 \log P + (-2.40 \pm 0.05)$.

§ $M_V = (0.22 \pm 0.22)([\text{Fe/H}] + 1.5) + (0.66 \pm 0.11)$

Table 1 lists these DCEPs and indicates the progress moving from the revised *Hipparcos* parallaxes to the *HST*-based parallaxes to *Gaia* (Gaia Collaboration et al. 2016b), in particular DR3 (GDR3; Gaia Collaboration et al. 2021). Table 2 shows the same for the RRL and T2Cs. The last column lists the goodness-of-fit parameter (GoF) which indicates the quality of the astrometric solution (in a way similar to the 'renormalised unit weight error', RUWE). The expectation value for the GoF is zero, with an error of unity.

Several things are noteworthy: (1) The improvement in accuracy from *HST*/FGS to GDR3 is about a factor of 10 for the RRL and T2Cs, but not really apparent for the DCEPs observed with the FGS. (2) For the DCEPs observed with *HST*/WFC3 and for the RRL/T2C, the GDR3 parallax errors are comparable, and the GoF values indicate good astrometric solutions (except for $\kappa$ Pav). (3) For most of the brighter DCEPs observed with *HST*/FGS the GoF values are poor, and those with GoF >8 would typically be excluded from any fitting. (4) The *HST* parallaxes are, on average, larger than the GDR3 ones. This is directly related to the parallax zero-point offset (PZPO; see below) in the *Gaia* data, and this allowed Groenewegen (2018b) to derive a global offset of $-49 \pm 18$ $\mu$as for DCEP in GDR2, based on nine DCEPs with GoF < 8—including RS Pup and its geometric distance from Kervella et al. (2014)—and non-*Gaia* parallax errors comparable to the GDR2 parallax errors.

The advent of the revised *Hipparcos* parallaxes and *HST*/FGS measurements led to new PL relations that were probably the standard ones until the arrival of *Gaia*. Fouqué et al. (2007) presented PL relations in seven bands, from *B* to *K*, and in two Wesenheit colours using 59 DCEPs with parallaxes based on the revised *Hipparcos* catalogue, *HST*/FGS, Baade–Wesselink techniques, and distances to open clusters (OCs) containing DCEPs.

Feast et al. (2008) used the revised *Hipparcos* parallaxes for the 142 RRL in the *Hipparcos* catalogue and the *HST*/FGS parallax of RR Lyrae (Benedict et al. 2002) to derive relations in the *V* and *K* band†. Benedict et al. (2011) presented PL relations based on the five RRL they observed with the FGS‡.

**Table 2.** RRL and T2C parallaxes (in mas) from *Hipparcos* to *Gaia*

| Name | Type | $\pi \pm \sigma_\pi$ (rev. *Hipparcos*) | $\pi \pm \sigma_\pi$ (*HST*) | $\pi \pm \sigma_\pi$ (GDR3) | GoF |
|------|------|------|------|------|------|
| RR Lyr | RRab | $3.46 \pm 0.64$ | $3.77 \pm 0.13$ | $3.985 \pm 0.027$ | 0.9 |
| SU Dra | RRab | $0.20 \pm 1.13$ | $1.42 \pm 0.16$ | $1.332 \pm 0.014$ | 0.4 |
| UV Oct | RRab | $2.44 \pm 0.81$ | $1.71 \pm 0.10$ | $1.838 \pm 0.012$ | 0.0 |
| XZ Cyg | RRab | $2.29 \pm 0.84$ | $1.67 \pm 0.17$ | $1.586 \pm 0.015$ | 4.1 |
| RZ Cep | RRc | $0.59 \pm 1.48$ | $2.54 \pm 0.19$ | $2.401 \pm 0.012$ | $-0.3$ |
| | | | | | |
| VY Pyx | BL Her | $5.01 \pm 0.44$ | $6.44 \pm 0.23$ | $3.950 \pm 0.019$ | $-3.6$ |
| $\kappa$ Pav | W Vir | $6.52 \pm 0.77$ | $5.57 \pm 0.28$ | $5.245 \pm 0.122$ | 38 |

Col. 2: Pulsation Type. Col. 3: Revised *Hipparcos* parallaxes (van Leeuwen 2007, 2008). Col. 4: *HST* parallaxes from Benedict et al. (2011). Col. 5: GDR3 parallaxes, with the GoF parameter in Col. 6.

## 2. The *Gaia* DR3 era

Numerous studies have been published related to PL relations of variable stars since the first *Gaia* data release. One recurrent issue has been the fact that point sources at, essentially, infinity (like quasi-stellar objects; QSOs) do not have, on average, a *Gaia* parallax of zero;

† $M_V = 0.214([\text{Fe/H}] + 1.38) + (0.54 \pm 0.10)$ and $M_K = -2.41(\log P + 0.252) + (-0.63 \pm 0.10)$, for fixed slopes.

‡ $M_V = 0.214([\text{Fe/H}] + 1.5) + (0.45 \pm 0.05)$ and, $M_K = -2.38(\log P + 0.28) + (-0.57 \pm 0.03)$ [solution 2 in their table 9], for fixed slopes.

there is a PZPO. For the QSOs this global offset was PZPO = −29 $\mu$as in DR2 (Lindegren et al. 2018) and −17 $\mu$as in DR3 (Lindegren et al. 2021a), in the sense $\pi_{true} = \pi_{Gaia}$ − PZPO.

After the publication of GDR2, numerous studies appeared that determined the PZPO using other classes of stars, typically at brighter magnitudes than the QSOs. Examples using GDR2 data for DCEPs are Riess et al. (2018b) and Groenewegen (2018b), who found a more negative value (−46 ± 13 $\mu$as and −49 ± 18 $\mu$as, respectively), and for RRL Muraveva et al. (2018) and Layden et al. (2019) (∼ −56 mas and −42 ± 13 mas, respectively). Many other classes of objects have been used to study the PZPO, like red clump stars, (detached) eclipsing binaries, red giant asteroseismic parallaxes, stars with parallaxes from Very Long Baseline Interferometry, W UMa-type binaries, and so on, and this continued when G(E)DR3 was released (Huang et al. 2021b; Stassun and Torres 2021; Ren et al. 2021; Vasiliev and Baumgardt 2021; Zinn 2021; Flynn et al. 2022; Maíz Apellániz 2022; Wang et al. 2022; Khan et al. 2023).

(Lindegren et al. 2021a, hereafter L21) provided a python script to the community which returned the PZPO (without an error bar) as a function of input parameters, namely ecliptic latitude ($\beta$), $G$-band magnitude, the `astrometric_params_solved` parameter, and either the effective wavenumber of the source used in the astrometric solution ($\nu_{eff}$, `nu_eff_used_in_astrometry` for the five-parameter solution `astrometric_params_solved` = 31) or the astrometrically estimated pseudo-colour of the source (`pseudocolour`) for the six-parameter solution (`astrometric_params_solved` = 95). The module is defined in the range $6 < G < 21$ mag, $1.24 < \nu_{eff} < 1.72 \mu m^{-1}$, corresponding to about $0.15 < (Bp - Rp) <$ 3.0 mag where $G$, $Bp$, and $Rp$ are the magnitudes in the *Gaia G*, BP, and RP band, respectively. (Maíz Apellániz 2022, hereafter MA22) provided an IDL script to determine the PZPO (without an error bar) using the same mathematical framework as in L21, but with updated coefficients.

Groenewegen (2021), hereafter G21, took a different approach, in particular regarding the spatial component of the PZPO. While in L21 and MA22 the spatial component is parameterised using a second-order polynomial fit, G21 determined the spatial component (for a fiducial magnitude of $G = 20$ mag) based on the HEALPix formalism (Górski et al. 2005)† The magnitude-dependent component is a combination of linear functions and based on wide binary systems, as in L21. Groenewegen (2021) found no real colour dependence of the PZPO.

Claims have been made that the L21 correction overcorrects the parallax offset (there are also claims to the contrary; see below). In some papers, one therefore considers a correction to the PZPO recipes, like

$$\pi_{true} = \pi_{Gaia} - \text{PZPO} + \Delta\pi \qquad (2)$$

(note that in some papers the sign of the last term is reversed).

Some papers do not use the (corrected) parallax but rather the Bayesian estimate of the distance from Bailer-Jones et al. (2021). This latter work includes the L21 correction and uses a prior on the distance. This prior is pre-calculated based on a mock catalogue of GDR3 data for HEALPix level 5. This prior may not be optimal for specific populations, like the Cepheids that mainly reside in the Galactic disc, or for RR Lyrae that are mainly found in the halo.

### 2.1. *Recent PL relations.*

Table 3 lists recent PL relations for DCEPs. Always consult the original reference, since typically additional PL relations are also given, e.g. in other photometric bands, fundamental (FU) mode versus first overtone (FO) versus FU+FO pulsators. After each reference some

---

† The HEALPix formalism is a convenient way to divide the sky into equal-area pixels. At HEALPix level 0 there are 12 pixels, and this increases by a factor of four for every next higher level. The HEALPix formalism is used by the *Gaia* team and is encoded in the `source_id` as pixel number = `source_id`/($2^{35} \times 4^{(12-\text{level})}$) for a given HEALPix level.

**Table 3.** Recent PL relations for DCEPs.

| $\alpha^1$ | $\beta$ | $\gamma^1$ | Band | GAL | $N$ | $\sigma$ | Remarks[2] |
|---|---|---|---|---|---|---|---|
| | | | Ripepi et al. (2023) | | | | |
| $-2.793 \pm 0.015$ | $17.333 \pm 0.010$ | $0.0$ | $G$ | LMC | 2477 | 0.169 | FU |
| $-2.830 \pm 0.033$ | $17.757 \pm 0.026$ | $0.0$ | $G$ | SMC | 843 | 0.251 | $P > 2.95$ d |
| $-3.317 \pm 0.007$ | $15.998 \pm 0.005$ | $0.0$ | $W_G$ | LMC | 2477 | 0.075 | |
| $-3.382 \pm 0.021$ | $16.592 \pm 0.026$ | $0.0$ | $W_G$ | SMC | 839 | 0.156 | $P > 2.95$ d |
| | | Ripepi et al. (2022); DR3, L21 + Riess et al. (2021a) correction, $1.1\sigma_\pi$ | | | | | |
| $-3.294 \pm 0.040$ | $-6.042 \pm 0.013$ | $0.0$ | $W_G$ | MW | 372 | 0.017 | FU/ABL |
| $-3.178 \pm 0.048$ | $-5.971 \pm 0.017$ | $-0.661 \pm 0.077$ | $W_G$ | MW | 372 | 0.016 | |
| $-3.176 \pm 0.044$ | $-5.988 \pm 0.018$ | $-0.520 \pm 0.090$ | $W_G$ | MW | 435 | 0.015 | F+1O 'best' |
| | | Breuval et al. (2022); DR3, BJ21 + correction (L21 + Riess et al. (2021a) correction) | | | | | |
| $-2.715F$ | $-4.153 \pm 0.060$ | | $V$ | MW | 183 | 0.22 | |
| $-2.715F$ | $-4.129 \pm 0.046$ | $-0.311 \pm 0.082$ | $V$ | MCs/MW | 2083 | | |
| $-2.816F$ | $-4.468 \pm 0.064$ | | $G$ | MW | 446 | 0.49 | |
| $-2.816F$ | $-4.384 \pm 0.047$ | $-0.462 \pm 0.089$ | $G$ | MCs/MW | 2083 | | |
| $-3.222F$ | $-5.899 \pm 0.033$ | | $K$ | MW | 65 | 0.17 | |
| $-3.222F$ | $-5.827 \pm 0.034$ | $-0.321 \pm 0.068$ | $K$ | MCs/MW | 2017 | | |
| $-3.324F$ | $-5.965 \pm 0.032$ | | [3.6] | MW | 21 | 0.20 | |
| $-3.324F$ | $-5.912 \pm 0.031$ | $-0.292 \pm 0.057$ | [3.6] | MCs/MW | 110 | | |
| $-3.338F$ | $-6.006 \pm 0.021$ | | $W_G$ | MW | 596 | 0.32 | |
| $-3.338F$ | $-5.959 \pm 0.025$ | $-0.384 \pm 0.051$ | $W_G$ | MCs/MW | 2473 | | |
| $-3.291F$ | $-5.882 \pm 0.038$ | | $W_{VI}$ | MW | 157 | 0.18 | |
| $-3.291F$ | $-5.851 \pm 0.035$ | $-0.201 \pm 0.071$ | $W_{VI}$ | MCs/MW | 2051 | | |
| $-3.323F$ | $-6.185 \pm 0.038$ | | $W_{JK}$ | MW | 63 | 0.18 | |
| $-3.323F$ | $-6.134 \pm 0.042$ | $-0.322 \pm 0.079$ | $W_{JK}$ | MCs/MW | 2014 | | |
| $-3.305F$ | $-5.955 \pm 0.024$ | | $W_H$ | MW | 60 | 0.16 | |
| $-3.305F$ | $-5.931 \pm 0.027$ | $-0.280 \pm 0.078$ | $W_H$ | LMC/MW | 130 | | |
| | | Owens et al. (2022); DR3, BJ21 + offset | | | | | |
| $-3.265F$ | $-5.693 \pm 0.023$ | | $K$ | MW | 37 | 0.142 | +18 F 5–60 d |
| $-3.284F$ | $-5.793 \pm 0.022$ | | [3.6] | MW | 37 | 0.087 | +18 F |
| $-3.401 \pm 0.091$ | $-5.986 \pm 0.023$ | | $W_{JK}$ | MW | 37 | 0.140 | +18 F |
| $-3.339 \pm 0.099$ | $-6.009 \pm 0.025$ | | $W_{VI}$ | MW | 37 | 0.151 | +18 F |
| | | Molinaro et al. (2023); DR3, L21 + fitted offset | | | | | |
| $-3.167 \pm 0.053$ | $-5.887 \pm 0.035$ | $-0.43 \pm 0.12$ | $K$ | MW | 443 | 0.017 | $-24.8 \pm 4.6$ |
| $-3.333 \pm 0.057$ | $-6.178 \pm 0.038$ | $-0.40 \pm 0.12$ | $W_{JK}$ | MW | 443 | 0.016 | $-21.9 \pm 5.0$ |
| $-3.191 \pm 0.042$ | $-6.018 \pm 0.042$ | $-0.32 \pm 0.10$ | $W_H(T)$ | MW | 430 | 0.016 | $-23.6 \pm 4.1$ |
| | | Riess et al. (2022); DR3, L21 + fitted offset, MW cluster Cepheids | | | | | |
| $-3.299F$ | $-5.902 \pm 0.026$ | $-0.217F$ | $W_H$ | MW | 17 | 0.06 | $-3 \pm 4$, G+S |
| $-3.299F$ | $-5.890 \pm 0.018$ | $-0.217F$ | $W_H$ | MW | 17 | 0.06 | $0F$ |
| $-3.36 \pm 0.07$ | $-5.893 \pm 0.018$ | $-0.217F$ | $W_H$ | MW | 17 | 0.06 | $0F$ |
| | | Cruz Reyes and Anderson (2023); DR3, L21 + fitted offset, MW cluster+field Cepheids | | | | | |
| $-3.299F$ | $-5.914 \pm 0.017$ | $-0.217F$ | $W_H$ | MW | 82 | | $-13 \pm 5$ |
| $-3.383 \pm 0.052$ | $-5.930 \pm 0.020$ | $-0.217F$ | $W_H$ | MW | 82 | | $-17 \pm 5$ |
| | | Riess et al. (2021a); DR3, L21 + fitted offset, MW field Cepheids | | | | | |
| $-3.28 \pm 0.06$ | $-5.915 \pm 0.030$ | $-0.20 \pm 0.13$ | $W_H$ | MW | 75 | | $-14 \pm 6$ |
| $-3.26F$ | $-5.915 \pm 0.022$ | $-0.17F$ | $W_H$ | MW | 75 | | $-14 \pm 6$ |
| $-3.26F$ | $-5.865 \pm 0.013$ | $-0.17F$ | $W_H$ | MW | 75 | | $0F$ |

keywords are listed as to how the distances or parallaxes were obtained. In most cases, this is based on GDR3 data and application of the L21 correction to individual objects. Sometimes an additional constant offset is fitted, or one used distances from BJ21. Always consult the original reference for details on the methodology.

PL relations are parameterised in the literature typically as $M = \alpha(\log P - p1) + \beta + \gamma([\text{Fe/H}] - p2)$, where $p1$ and $p2$ are constants. Zero points $\beta$ in the table have been rescaled to a period of 10 days and [Fe/H] = 0 dex where appropriate. Wesenheit functions are defined as follows: $W_G = G - 1.90(Bp - Rp)$; $W_{VI} = I - \varepsilon(V - I)$, with $\varepsilon = 1.387$ (Breuval et al. 2022)

**Table 3.** Recent PL relations for DCEPs (Continued) and T2C.

| $\alpha^1$ | $\beta$ | $\gamma^1$ | Band | GAL | $N$ | $\sigma$ | Remarks[2] |
|---|---|---|---|---|---|---|---|
| | | Lin et al. (2022); DR3, L21, MW cluster Cepheids | | | | | |
| $-2.94 \pm 0.12$ | $-5.87 \pm 0.11$ | | $W_G$ | MW OC | 46 | 0.25 | |
| | Wang et al. (2018); distances from $JHK$ PL relations of 31 DCEP in OCs (Chen et al. 2017) | | | | | | |
| $-3.248 \pm 0.018$ | $-5.781 \pm 0.017$ | | $W1$ | MW | 255 | 0.081 | FU |
| $-3.266 \pm 0.027$ | $-5.811 \pm 0.027$ | | $W2$ | MW | 190 | 0.107 | FU |
| $-3.298 \pm 0.024$ | $-5.765 \pm 0.024$ | | [3.6] | MW | 85 | 0.066 | FU |
| $-3.245 \pm 0.024$ | $-5.747 \pm 0.027$ | | [4.5] | MW | 99 | 0.073 | FU |
| | | Lemasle et al. (2022) | | | | | |
| $-2.436 \pm 0.013$ | $-5.632 \pm 0.019$ | | $W_{W1W2}$ | LMC | 2326 | 0.149 | FU |
| | | | T2C[3] | | | | |
| | | Ripepi et al. (2023) | | | | | |
| $-1.905 \pm 0.036$ | $18.731 \pm 0.033$ | | $G$ | LMC | 205 | 0.275 | |
| $-2.140 \pm 0.100$ | $19.105 \pm 0.098$ | | $G$ | SMC | 42 | 0.372 | |
| $-2.577 \pm 0.019$ | $17.516 \pm 0.033$ | | $W_G$ | LMC | 197 | 0.138 | |
| $-2.505 \pm 0.054$ | $17.843 \pm 0.052$ | | $W_G$ | SMC | 42 | 0.190 | |
| | Ngeow et al. (2022b); distances GCC from Baumgardt and Vasiliev (2021) | | | | | | |
| $-2.41 \pm 0.03$ | $-1.09 \pm 0.03$ | | $K$ | GCC | 48 | 0.10 | |
| $-2.46 \pm 0.03$ | $-1.28 \pm 0.03$ | | $W_{JK}$ | GCC | 46 | 0.08 | |
| | Wielgórski et al. (2022); DR3 + L21 (also for G21) / LMC | | | | | | |
| $-2.387 \pm 0.030$ | $-1.013 \pm 0.011$ | | $K$ | LMC | 62 | 0.087 | BLH+WVIR |
| $-2.544 \pm 0.029$ | $-1.166 \pm 0.011$ | | $W_{JK}$ | LMC | 61 | 0.087 | BLH+WVIR |
| $-2.225 \pm 0.089$ | $-1.169 \pm 0.024$ | | $K$ | MW | 14 | 0.065 | BLH+WVIR |
| $-2.399 \pm 0.095$ | $-1.355 \pm 0.022$ | | $W_{JK}$ | MW | 15 | 0.061 | BLH+WVIR |

Notes. [1] $F$ stands for a parameter that was fixed. [2] Any parallax offsets quoted in Col. 8 are in $\mu$as.
[3] For T2C zero points are at $P = 1$ day.

or 1.61 (Owens et al. 2022); $W_{JK} = K - \varepsilon(J - K)$, with $\varepsilon = 0.735$ (Breuval et al. 2022), 0.71 (Owens et al. 2022), 0.69 (Molinaro et al. 2023; Wielgórski et al. 2022), or 0.618 (Ngeow et al. 2022b); $W_{W1W2} = W2 - 2.0(W1 - W2)$; and $W_H = $ F160W $- 0.386$(F555W $-$ F814W). $W_H(T)$ means that $VIH$-band photometry was transformed to the $HST$ system. In particular for $W_{VI}$ and $W_{JK}$, different functional forms are adopted in the literature that make a direct comparison of PL relations quasi-impossible. The different coefficients relate to differences in the adopted reddening laws (see Section 2.3).

One of the lingering open questions regarding DCEPs was the metallicity dependence of the PL relation. Significant progress has been made over the past years. In their derivation of the metallicity term Breuval et al. (2022) adopt a single metallicity for the LMC DCEPs ($-0.409$ dex from Romaniello et al. 2022, with an uncertainty of 0.020 dex) and SMC DCEPs ($-0.75 \pm 0.05$ dex from Gieren et al. 2018), but also for the Milky Way (MW) a single metallicity was adopted, typically ranging from ($+0.087 \pm 0.056$) dex to ($+0.099 \pm 0.089$) dex from the $V$ to $K$ bands and the Wesenheit indices, depending on the exact sample per filter)†. The approximation of a single metallicity for the MW sample does not make use of the actual range in metallicities that MW DCEPs cover.

Ripepi et al. (2022) and Molinaro et al. (2023) use exclusively MW DCEPs to constrain the metallicity term. The full range in [Fe/H] is from $-0.83$ to $+0.55$ dex (median = +0.05 dex, dispersion = 0.07 dex, from the 499 stars used by Ripepi et al. 2022). These two papers find steeper slopes than Breuval et al. (2022) but with larger error bars, and, formally, the slopes agree at the $2\sigma$ level. One should also keep in mind that, in general, the adopted or derived PZPO correlate with the coefficients $\alpha$, $\beta$, and $\gamma$ (see, e.g., table 7 of Groenewegen 2018b or table 7 of Breuval et al. 2020 based on GDR2. Parallaxes are intrinsically more accurate in GDR3, but this correlation should still hold to some extent also for GDR3 data).

† For comparison, Gieren et al. (2018) adopted [Fe/H] = $-0.34 \pm 0.06$ dex for the LMC and [Fe/H] = $+0.00 \pm 0.05$ dex for the MW.

Grosso modo, Pl relations derived by different authors appear to be in reasonably good agreement. A noticeable exception are the differences between Breuval et al. (2022) and Owens et al. (2022), in particular the additional PZPO that has a different sign; but see footnote 8 in Breuval et al. (2022) for an explanation that is related to the metallicity term.

Table 4 lists recent PL relations for RRL. Zero points $\beta$ have been rescaled to a period of one day and [Fe/H] = 0 dex where appropriate. Coefficients of 0.69 and 1.90 were adopted for the $W_{JK}$ and $W_G$ relations, unless noted otherwise. From a procedural point of view, the work of Mullen et al. (2023) is interesting as it takes into account that the errors in the astrometric solution are underestimated, and they use Fabricius et al. (2021) to correct their results (see also Ripepi et al. 2022, who apply a correction to their Cepheid sample). Li et al. (2023) adopt a Bayesian estimate to estimate the distance from the parallax, although not from the work of Bailer-Jones et al. (2021); instead, they use a prior that is more appropriate for RRL, namely a power law in Galactocentric distance (Huang et al. 2021a).

**Table 4.** Recent PL relations for RRL.

| $\alpha$ | $\beta$ | $\gamma$ | Band | GAL | $N$ | $\sigma$ | Remarks[2] |
|---|---|---|---|---|---|---|---|
| \multicolumn Bhardwaj et al. (2023); DR3, L21 | | | | | | | |
| $-2.37 \pm 0.02$ | $-0.80 \pm 0.03$ | $0.18 \pm 0.01$ | $K$ | MW/GC | 1077 | 0.05 | RRabcd |
| $-2.73 \pm 0.02$ | $-1.06 \pm 0.03$ | $0.16 \pm 0.02$ | $W_{JK}$ | MW/GC | 1096 | 0.06 | |
| Zgirski et al. (2023); DR3, BJ21 | | | | | | | |
| $-3.03 \pm 0.25$ | $-1.06 \pm 0.04$ | $0.083 \pm 0.025$ | $K$ | MW | 28 | 0.073 | RRabc |
| $-3.26 \pm 0.23$ | $-0.80 \pm 0.03$ | $0.083 \pm 0.024$ | $W_{JK}$ | MW | 28 | 0.070 | |
| Looijmans et al. (2023); DR3, fit global offset | | | | | | | |
| $+0.393 \pm 0.049$ | $+0.76 \pm 0.05$ | $0.273 \pm 0.022$ | $V$ | MW/GC | 200 | | $-14 \pm 9$ |
| $-2.45 \pm 0.11$ | $-0.86 \pm 0.04$ | $0.171 \pm 0.018$ | $K$ | MW/GC | 200 | | $-15 \pm 5$ |
| $-2.48 \pm 0.07$ | $-0.93 \pm 0.02$ | $0.121 \pm 0.010$ | $W_G$[1] | MW/GC | 200 | | $-8 \pm 3$ |
| Mullen et al. (2023); DR3, L21, error inflation factor | | | | | | | |
| $-2.44 \pm 0.10$ | $-0.841 \pm 0.008$ | $0.144 \pm 0.014$ | $W1$ | MW/GC | 1052 | 0.02 | |
| $-2.54 \pm 0.10$ | $-0.857 \pm 0.009$ | $0.151 \pm 0.014$ | $W2$ | MW/GC | 397 | 0.02 | |
| Li et al. (2023); DR3, L21 + Bayesian estimate | | | | | | | |
| | $+1.106 \pm 0.021$ | $0.350 \pm 0.016$ | $G$ | MW | 205 | 0.12 | RRab |
| $-2.465 \pm 0.084$ | $-0.792 \pm 0.043$ | $0.161 \pm 0.011$ | $K$ | MW | 159 | 0.14 | |
| $-2.452 \pm 0.080$ | $-0.834 \pm 0.031$ | $0.179 \pm 0.011$ | $W1$ | MW | 164 | 0.09 | |
| Garofalo et al. (2022); DR3, fit global offset | | | | | | | |
| | $+1.13 \pm 0.03$ | $0.33 \pm 0.02$ | $V$ | MW | 270 | 0.02 | $-33 \pm 2$ |
| $-2.49 \pm 0.21$ | $-0.88 \pm 0.09$ | $0.14 \pm 0.03$ | $W_G$ | MW | 190 | 0.09 | $-33 \pm 3$ |
| Gilligan et al. (2021); DR3, L21 + fit global offset, error inflation | | | | | | | |
| $-2.78 \pm 0.12$ | $-1.018 \pm 0.020$ | $0.115 \pm 0.016$ | $W1$ | MW | 150 | | $10 \pm 7$ |
| Ngeow et al. (2022a); GCC (DR3) | | | | | | | |
| $-1.43 \pm 0.14$ | $+0.243 \pm 0.055$ | $0.144 \pm 0.018$ | $i$ | GCC | 321 | 0.15 | RR0 (= FU) |
| $-2.60 \pm 0.13$ | $+0.093 \pm 0.051$ | $0.193 \pm 0.017$ | $W_{ri}$[3] | GCC | 325 | 0.15 | |
| Neeley et al. (2019); DR2, PZPO = $-30$ mas | | | | | | | |
| $-2.40 \pm 0.27$ | $-0.87 \pm 0.02$ | $0.18 \pm 0.03$ | [3.6] | MW | 55 | 0.18 | |
| $-2.45 \pm 0.28$ | $-0.89 \pm 0.02$ | $0.18 \pm 0.03$ | [4.5] | MW | 55 | 0.18 | |
| $-2.60 \pm 0.25$ | $-1.01 \pm 0.02$ | $0.13 \pm 0.04$ | $W_{VI}$[4] | MW | 55 | 0.18 | |

Notes. [1] They used $W_G = G - 1.85(Bp - Rp)$. [2] Any parallax offsets quoted in Col. 8 are in $\mu$as.
[3] $W_{ri} = r - 4.051(r - i)$. [4] $W_{VI} = I - 1.467(V - I)$.

## 2.2. *Cepheids in Open Clusters*

Since the release of GDR2 there has been a revolution in the census and study of OCs; see Hunt and Reffert (2023) and references therein. Not surprisingly, there has been renewed interest specifically for DCEPs in OCs as well, extending pre-*Gaia* work by, e.g., Anderson et al. (2013). Recent papers on the subject are by Medina et al. (2021), Zhou and Chen (2021), Lin et al. (2022), Riess et al. (2022), Hao et al. (2022), and Cruz Reyes and Anderson (2023). In addition some new results are presented in this work.

Table 5 contains the derived parallaxes and errors of the clusters containing DCEPs, where the recent work by Cruz Reyes and Anderson (2023) has been taken as reference for the list of clusters. Riess et al. (2022) considered clusters for which *HST* photometry is available. The difference between Riess et al. (2022) and Cruz Reyes and Anderson (2023) is in the membership list. Riess et al. (2022) use cluster members derived from GDR2 data Cantat-Gaudin et al. (2018), although parallaxes from GDR3 were used in the distance determination, while Cruz Reyes and Anderson (2023) independently determined cluster members from GDR3 data. There are also small differences in selecting the final sample, e.g., Cruz Reyes and Anderson (2023) only consider members in the range $12.5 < G < 17$ mag and $0.23 < Bp - Rp < 2.75$ mag, where the L21 correction results in the smallest residuals. Indeed, both works include the L21 correction. The differences in parallaxes are small (Col. $3 - $ Col. $2 = 5.7 \pm 2.0$ $\mu$as for 15 clusters). Column 4 gives the parallaxes as quoted by Hao et al. (2022). They do not include a PZPO.

Column 5 lists the results obtained by me using the membership list of Cruz Reyes and Anderson (2023), and they do not include a PZPO either. No selection based on magnitude or colour was made; however a selection based on the GoF as well as $4\sigma$ clipping on parallaxes and proper motions (PMs) in Right Ascension and Declination† relative to the median values and using ($1.486 \times$ median-absolute-deviation) as an estimate for $\sigma$ was performed. These two estimates agree very well with (Col. $5 - $ Col. $4$) $= 0.8 \pm 2.5$ $\mu$as for 20 clusters.

A few remaining remarks on the results in Table 5: First, the difference between Cols. 2/3 and Cols. 4/5 is the inclusion of the L21 correction. The average PZPO (Col. $5 - $ Col. $3$) $= -35.2 \pm 2.0$ $\mu$as for 30 clusters. Second, Cepheids can be assigned to *different* clusters by different authors. This is the meaning of the pair of curly brackets at the end of the table. Membership is typically based on position, parallax and PM in contemporary searches. Typically, there is not enough accurate (*Gaia*) radial velocity (RV) data available to be used as an additional criterion. Third, one notices that there is a lower limit to the error in the mean cluster parallax of 6–8 $\mu$as in Riess et al. (2022) and Cruz Reyes and Anderson (2023) and 10–12 $\mu$as in the present work.

The total parallax uncertainty adds in quadrature the statistical uncertainty determined as the error on the median‡, the uncertainty of where the Cepheid is located in the cluster§, and the systematic contribution due to the angular covariance of parallaxes at small scales. The first term is typically small, as this goes as $1/\sqrt{N}$. When replacing the parallax of the Cepheid with the parallax of the cluster, there is an uncertainty of where the Cepheid is located with respect to the centre of the cluster. To estimate this effect, the radius on the sky containing 68% of cluster members is converted into a distance. This effect is also small in most cases. The total error is dominated by the last term. Parallaxes are correlated on all spatial scales, but the effect is largest on small scales. Riess et al. (2022) and Cruz Reyes and Anderson (2023) largely follow Maíz Apellániz et al. (2021). There are other studies on the matter (e.g., Lindegren et al. 2021b), and I followed Vasiliev and Baumgardt (2021, their Eq. 1). A better understanding of the behaviour of the spatial covariance and smaller values are expected in further *Gaia* releases.

† Cruz Reyes and Anderson (2023) use $6\sigma$ clipping.

‡ Cruz Reyes and Anderson (2023) use the weighted mean.

§ This effect is not considered in either Riess et al. (2022) or Cruz Reyes and Anderson (2023).

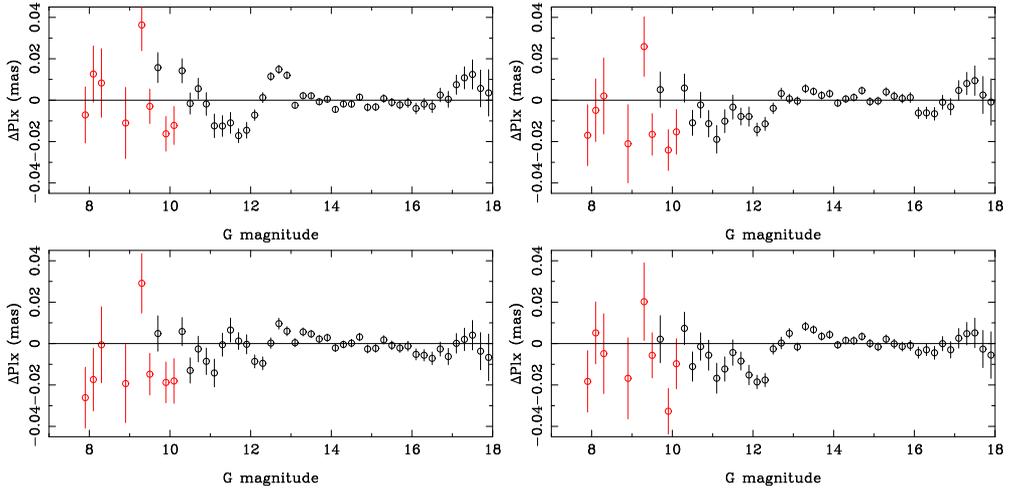

**Figure 1.** Residuals between individual and cluster parallaxes summed over all clusters and plotted against *G* magnitude. Bins plotted in red contain less than 0.1% of all stars. PZPO models are: no correction (top left), L21 (top right), MA22 (bottom left), G21 (bottom right).

Studies have typically used the L21 correction when considering the PZPO. Table 6 lists results using other recipes. Columns 2 and 6 have been copied from Table 5. They reflect the results from Cruz Reyes and Anderson (2023) (which includes L21) and this work (without PZPO). Columns 3–5 are from this work using the recipe from L21, MA22, and G21 for HEALPix level 2.

The differences between Columns 2 and 3 are minor (Col. 3 − Col. 2 = $-0.3 \pm 2.0 \, \mu$as for 30 clusters). They are solely related to minor differences in the selection of the final sample (differences in selection of magnitude and colour, differences in the clipping in parallaxes and PMs, etc.), since both sets use the L21 correction. The differences between using MA22 or L21 is also small (Col. 4 − Col. 3 = $-1.3 \pm 2.0 \, \mu$as). The differences between using G21 and L21 is larger: (Col. 5 − Col. 3) = $+1.9 \pm 2.0 \, \mu$as. Using no PZPO gives (Col. 6 − Col. 3) = $-35.0 \pm 2.0 \, \mu$as.

Figure 1 provides another view of these differences (similar to figure 2 in Cruz Reyes and Anderson 2023). It shows the difference between individual parallaxes and the mean cluster parallax for all cluster members summed over all clusters as a function of *G* magnitude. Plotted are the models with no correction (top left), L21 (top right), MA22 (bottom left), and G21 (bottom right). The model without PZPO shows the characteristic behaviour as a function of magnitude; see figure 20 in L21 (or figure 7 in G21 or figure 8 in MA22).

All three correction schemes reduce the scatter, in particular the trends between about 11.5 and 13th magnitude, but none does so in a perfect manner. Some residual trends and offsets remain. Table 7 lists the r.m.s. values in different magnitude bins and overall. It is difficult to draw firm conclusions, although the 11.5–13 magnitude range remains problematic even after applying any of the PZPO corrections, at least for the sample studied here.

Figure 2 shows the same plotted against $(Bp - Rp)$ colour but only for the no-PZPO and the L21 correction as the other two corrections are very similar. Although the L21 correction includes a dependence on colour, the difference with no correction seems small. The offsets for the bluest and reddest colours appear real (see the discussion in Cruz Reyes and Anderson 2023)). L21 recommend not to use their correction outside the range corresponding to 0.15–3.0 mag in $(Bp - Rp)$.

**Table 5.** Clusters containing Cepheids

| Cluster | $\pi \pm \sigma_\pi$ (1) | $\pi \pm \sigma_\pi$ (2) | $\pi \pm \sigma_\pi$ (3) | $\pi \pm \sigma_\pi$ (4) | $N$ | $R$ (') |
|---|---|---|---|---|---|---|
| Berkeley 58 | $336 \pm 8$ | $336 \pm 7$ | $290 \pm 20$ | $302 \pm 11$ | 85 | 4.0 |
| NGC 129 | $559 \pm 7$ | $557 \pm 7$ | $530 \pm 30$ | $529 \pm 10$ | 290 | 6.5 |
| FSR 0951 | $594 \pm 7$ | $610 \pm 7$ | $570 \pm 30$ | $566 \pm 11$ | 128 | 6.8 |
| vandenBergh 1 | $579 \pm 10$ | $585 \pm 10$ | $550 \pm 40$ | $543 \pm 13$ | 50 | 1.9 |
| Ruprecht 79 | $274 \pm 8$ | $281 \pm 7$ | $240 \pm 20$ | $244 \pm 11$ | 99 | 2.8 |
| NGC 5662 | $1322 \pm 7$ | $1336 \pm 6$ | $1300 \pm 30$ | $1292 \pm 12$ | 225 | 18.1 |
| Lynga 6 | $408 \pm 9$ | $421 \pm 8$ | $380 \pm 40$ | $383 \pm 11$ | 157 | 5.2 |
| NGC 6067 | $496 \pm 7$ | $513 \pm 7$ | $470 \pm 30$ | $473 \pm 11$ | 671 | 4.5 |
| NGC 6087 | $1057 \pm 7$ | $1073 \pm 7$ | $1020 \pm 30$ | $1025 \pm 11$ | 154 | 7.7 |
| IC 4725 | $1540 \pm 7$ | $1554 \pm 6$ | $1520 \pm 50$ | $1514 \pm 12$ | 325 | 11.6 |
| NGC 6649 | $508 \pm 8$ | $514 \pm 7$ | $470 \pm 50$ | $474 \pm 11$ | 301 | 3.1 |
| UBC 129 | $886 \pm 7$ | $880 \pm 7$ | $850 \pm 20$ | $852 \pm 11$ | 116 | 11.7 |
| UBC 130 | $428 \pm 8$ | $425 \pm 9$ | $400 \pm 20$ | $396 \pm 11$ | 36 | 2.4 |
| NGC 7790 | $331 \pm 8$ | $322 \pm 7$ | $290 \pm 30$ | $293 \pm 11$ | 114 | 3.0 |
| Czernik 41 | | $407 \pm 8$ | $370 \pm 40$ | $373 \pm 11$ | 103 | 2.7 |
| FSR 0384 | | $520 \pm 8$ | | $492 \pm 12$ | 56 | 3.3 |
| NGC 103 | | $317 \pm 7$ | $280 \pm 20$ | $284 \pm 11$ | 104 | 3.1 |
| NGC 6664 | | $504 \pm 7$ | $460 \pm 40$ | $469 \pm 11$ | 286 | 6.3 |
| UBC 106 | | $443 \pm 7$ | $410 \pm 30$ | $407 \pm 10$ | 328 | 6.8 |
| UBC 290 | | $639 \pm 6$ | $610 \pm 20$ | $604 \pm 11$ | 204 | 9.9 |
| UBC 375 | | $562 \pm 7$ | | $536 \pm 12$ | 142 | 8.3 |
| UBC 533 | | $878 \pm 9$ | | $834 \pm 12$ | 52 | 6.0 |
| CWNU 0175 | | $732 \pm 9$ | | $709 \pm 12$ | 31 | 7.0 |
| Cl XPup | | $363 \pm 6$ | | $330 \pm 10$ | 122 | 16.2 |
| Cl IQNor | | $544 \pm 9$ | | $504 \pm 12$ | 42 | 7.7 |
| Cl STTau | | $953 \pm 8$ | | $921 \pm 11$ | 66 | 9.9 |
| Cl SXVel | | $497 \pm 7$ | | $459 \pm 10$ | 73 | 12.6 |
| Cl V378Cen | | $518 \pm 8$ | | $485 \pm 11$ | 95 | 6.3 |
| ⎰ UBC 231 | $356 \pm 8$ | $345 \pm 8$ | | $312 \pm 11$ | 56 | 7.9 |
| ⎱ OC − 0717 | | | $370 \pm 10$ | $370 \pm 10$ | 64 | 24.4 |
| ⎰ Ruprecht 93 | | $482 \pm 7$ | | $448 \pm 11$ | 169 | 4.7 |
| ⎱ NGC 3496 | $439 \pm 8$ | | | $408 \pm 10$ | 464 | 7.6 |

Notes. Parallaxes in $\mu$as. Col. 2: Riess et al. (2022); Col. 3: Cruz Reyes and Anderson (2023); Col. 4: Hao et al. (2022) (the original work quotes parallaxes in mas to two decimal places); Col. 5: This work. Columns 2 and 3 include the L21 PZPO correction; Cols. 4–5 include no PZPO. Col. 6: Number of stars entering the distance calculation in the present work. Col. 7: Size of the cluster in arcminutes defined as the radius that includes 68% of its members. Names in Col. 1 starting with "Cl" are new clusters proposed by Cruz Reyes and Anderson (2023).

## 2.3. *Reddening and the Wesenheit function*

Dereddening the observed magnitudes for the effect of dust extinction in the MW foreground or internal to a galaxy represents a crucial step in obtaining absolute calibrated magnitudes and PL relations. The Wesenheit function (Madore 1982) was introduced to eliminate the effects of reddening, but the use of different coefficients in the literature mentioned in Section 2.1 reminds us of the fact that any given coefficient is only valid for one particular reddening law. For a Wesenheit function $W = a - R(b - c)$, $R$ is given by $R = A_a/(A_b - A_c)$.

I consider here the reddening laws of Cardelli et al. (1989) with the update in the optical published by O'Donnell (1994, hereafter COD), Fitzpatrick (1999, hereafter F99)†, and Wang

---

† Using the routine https://idlastro.gsfc.nasa.gov/ftp/pro/astro/fm_unred.pro

**Table 6.** Comparing different PZPO models in clusters

| Cluster | $\pi \pm \sigma_\pi$ (1) | $\pi \pm \sigma_\pi$ (2) | $\pi \pm \sigma_\pi$ (3) | $\pi \pm \sigma_\pi$ (4) | $\pi \pm \sigma_\pi$ (5) |
|---|---|---|---|---|---|
| Berkeley 58 | $336 \pm 7$ | $330 \pm 11$ | $329 \pm 11$ | $338 \pm 11$ | $302 \pm 11$ |
| NGC 129 | $557 \pm 7$ | $558 \pm 10$ | $556 \pm 10$ | $564 \pm 10$ | $529 \pm 10$ |
| FSR 0951 | $610 \pm 7$ | $607 \pm 11$ | $607 \pm 11$ | $588 \pm 11$ | $566 \pm 11$ |
| vandenBergh 1 | $585 \pm 10$ | $585 \pm 12$ | $582 \pm 12$ | $576 \pm 13$ | $543 \pm 13$ |
| Ruprecht 79 | $281 \pm 7$ | $275 \pm 11$ | $274 \pm 11$ | $256 \pm 11$ | $244 \pm 11$ |
| NGC 5662 | $1336 \pm 6$ | $1328 \pm 12$ | $1329 \pm 12$ | $1358 \pm 12$ | $1292 \pm 12$ |
| Lynga 6 | $421 \pm 8$ | $420 \pm 11$ | $417 \pm 11$ | $400 \pm 11$ | $383 \pm 11$ |
| NGC 6067 | $513 \pm 7$ | $514 \pm 11$ | $513 \pm 11$ | $495 \pm 11$ | $473 \pm 11$ |
| NGC 6087 | $1073 \pm 7$ | $1065 \pm 11$ | $1066 \pm 11$ | $1065 \pm 11$ | $1025 \pm 11$ |
| IC 4725 | $1554 \pm 6$ | $1552 \pm 12$ | $1551 \pm 12$ | $1544 \pm 12$ | $1514 \pm 12$ |
| NGC 6649 | $514 \pm 7$ | $514 \pm 11$ | $511 \pm 11$ | $487 \pm 11$ | $474 \pm 11$ |
| UBC 129 | $880 \pm 7$ | $882 \pm 11$ | $880 \pm 11$ | $886 \pm 11$ | $852 \pm 11$ |
| UBC 130 | $425 \pm 9$ | $426 \pm 11$ | $423 \pm 12$ | $431 \pm 12$ | $396 \pm 11$ |
| NGC 7790 | $322 \pm 7$ | $321 \pm 11$ | $320 \pm 11$ | $341 \pm 11$ | $293 \pm 11$ |
| Czernik 41 | $407 \pm 8$ | $407 \pm 11$ | $404 \pm 11$ | $406 \pm 11$ | $373 \pm 11$ |
| FSR 0384 | $520 \pm 8$ | $520 \pm 12$ | $519 \pm 12$ | $541 \pm 11$ | $492 \pm 12$ |
| NGC 103 | $317 \pm 7$ | $312 \pm 11$ | $312 \pm 11$ | $320 \pm 11$ | $284 \pm 11$ |
| NGC 6664 | $504 \pm 7$ | $507 \pm 11$ | $504 \pm 11$ | $477 \pm 11$ | $469 \pm 11$ |
| UBC 106 | $443 \pm 7$ | $445 \pm 10$ | $443 \pm 10$ | $511 \pm 10$ | $407 \pm 10$ |
| UBC 290 | $639 \pm 6$ | $641 \pm 11$ | $641 \pm 11$ | $626 \pm 11$ | $604 \pm 11$ |
| UBC 375 | $562 \pm 7$ | $566 \pm 12$ | $563 \pm 11$ | $575 \pm 12$ | $536 \pm 12$ |
| UBC 533 | $878 \pm 9$ | $871 \pm 12$ | $870 \pm 12$ | $896 \pm 12$ | $834 \pm 12$ |
| CWNU 0175 | $732 \pm 9$ | $736 \pm 12$ | $734 \pm 12$ | $727 \pm 11$ | $709 \pm 12$ |
| Cl XPup | $363 \pm 6$ | $368 \pm 10$ | $366 \pm 10$ | $381 \pm 10$ | $330 \pm 10$ |
| Cl IQNor | $544 \pm 9$ | $542 \pm 12$ | $541 \pm 12$ | $571 \pm 12$ | $504 \pm 12$ |
| Cl STTau | $953 \pm 8$ | $962 \pm 12$ | $960 \pm 12$ | $935 \pm 12$ | $921 \pm 11$ |
| Cl SXVel | $497 \pm 7$ | $494 \pm 10$ | $495 \pm 10$ | $478 \pm 10$ | $459 \pm 10$ |
| Cl V378Cen | $518 \pm 8$ | $520 \pm 11$ | $519 \pm 11$ | $538 \pm 11$ | $485 \pm 11$ |
| UBC 231 | $345 \pm 8$ | $350 \pm 12$ | $349 \pm 12$ | $371 \pm 12$ | $312 \pm 11$ |
| Ruprecht 93 | $482 \pm 7$ | $484 \pm 11$ | $484 \pm 11$ | $473 \pm 11$ | $448 \pm 11$ |

Notes. Parallaxes in $\mu$as. Col. 2: Cruz Reyes and Anderson (2023); Cols. 3–6: This work, with the PZPOs from L21 (Col. 3), MA22 (Col. 4), G21 for HEALPix level 2 (Col. 5), no PZPO (Col. 6).

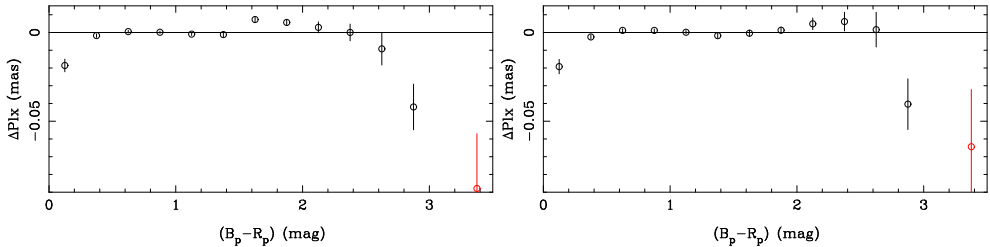

**Figure 2.** Residuals between individual and cluster parallaxes summed over all clusters and plotted against $(Bp - Rp)$ colour for no PZPO correction (left) and the L21 correction (right). Bins plotted in red contain less than 0.1% of all stars.

**Table 7.** Root-mean-square residuals between different PZPO models as a function of magnitude

| $G$ | no PZPO | L21 | MA22 | G21 |
|---|---|---|---|---|
| (mag) | ($\mu$as) | ($\mu$as) | ($\mu$as) | ($\mu$as) |
| All | 9.78 | 9.37 | 9.45 | 9.41 |
| $G \leq 13$ mag | 13.37 | 12.83 | 13.10 | 13.06 |
| $13 < G < 17$ mag | 2.35 | 3.48 | 3.60 | 3.39 |
| $17 \leq G \leq 18$ mag | 8.62 | 6.06 | 3.97 | 4.33 |

and Chen (2023, hereafter W23). The reddening laws are continuous functions of wavelength and, therefore, in order to calculate the coefficient of a Wesenheit function one has to adopt certain a wavelength for a certain filter. This characteristic wavelength depends on the under-lying spectrum of the star, but also on the reddening itself. I have calculated $\lambda_{\text{eff}}$ defined as

$$\lambda_{\text{eff}} = \frac{\int \lambda R(\lambda) M(\lambda) 10^{-0.4 A_\lambda} \, d\lambda}{\int R(\lambda) M(\lambda) 10^{-0.4 A_\lambda} \, d\lambda}, \tag{3}$$

where $R(\lambda)$ is the filter response curve, $M(\lambda)$ the flux of a MARCS model (Gustafsson et al. 2008), and $A_\lambda$ the reddening for the following filters: *Gaia G*, $Bp$, $Rp$, Johnson $V$, Cousins $I$ ($I_c$), OGLE-IV ($V_o$, $I_o$), 2MASS $JHK$, and *HST* F555W, F814W, and F160W.

I took 5000 K as a representative effective temperature of DCEPs, and 3600 K and 7000 K as extremes at the longest and shortest periods, respectively. The impact of choosing a metallicity and reddening was also investigated. The results are listed in Table 8. As expected, increased reddening and cooler stars lead to longer effective wavelengths. The effect of metallicity is negligible (at most 2 nm in the $Bp$ band). Overall, the effect is most noticeable in the broadband *Gaia G* filter and in the optical in general, and it essentially vanishes in the infrared (IR).

**Table 8.** Effective wavelengths (in nm) for in several filters under different conditions.

| Filter | Model 1 | 2 | 3 | 4 | 5 | 6 | 7 |
|---|---|---|---|---|---|---|---|
| $G$ | 702 | 769 | 658 | 701 | 700 | 669 | 736 |
| $Bp$ | 563 | 578 | 538 | 561 | 559 | 549 | 577 |
| $Rp$ | 794 | 826 | 785 | 795 | 796 | 781 | 811 |
| $V$ | 556 | 560 | 551 | 556 | 555 | 553 | 560 |
| $I_c$ | 790 | 795 | 788 | 790 | 790 | 787 | 793 |
| $V_o$ | 541 | 543 | 537 | 541 | 540 | 538 | 544 |
| $I_o$ | 798 | 803 | 796 | 799 | 799 | 795 | 802 |
| $J$ | 1237 | 1240 | 1231 | 1236 | 1236 | 1234 | 1238 |
| $H$ | 1643 | 1648 | 1641 | 1643 | 1643 | 1641 | 1645 |
| $K$ | 2153 | 2154 | 2153 | 2163 | 2153 | 2152 | 2154 |
| F555W | 545 | 548 | 540 | 544 | 544 | 541 | 549 |
| F814W | 814 | 828 | 810 | 814 | 814 | 807 | 823 |
| F160W | 1532 | 1541 | 1527 | 1532 | 1532 | 1531 | 1534 |

Standard is $\log g = 2.0$ and $M = 5$ M$_\odot$ for all MARCS models. Model 1: (default) $T_{\text{eff}} = 5000$ K, $[Z] = +0.0$, $E(B-V) = 0.5$ mag, Model 2: $T_{\text{eff}} = 3600$ K, $[Z] = +0.0$, $E(B-V) = 0.5$, Model 3: $T_{\text{eff}} = 7000$ K, $[Z] = +0.0$, $E(B-V) = 0.5$, Model 4: $T_{\text{eff}} = 5000$ K, $[Z] = -0.5$, $E(B-V) = 0.5$, Model 5: $T_{\text{eff}} = 5000$ K, $[Z] = -1.0$, $E(B-V) = 0.5$, Model 6: $T_{\text{eff}} = 5000$ K, $[Z] = +0.0$, $E(B-V) = 0.1$. Model 7: $T_{\text{eff}} = 5000$ K, $[Z] = +0.0$, $E(B-V) = 1.0$.

Table 9 lists the relative reddenings for the three different reddening laws (and the indicated values for $R_V$) for models 1/2/3 that cover the extreme range in changes in effective wavelength. The bottom part gives the relevant coefficient for different Wesenheit indices. The conclusion is that the coefficients vary considerably depending on the reddening law. In the case of the Wesenheit index based on the *Gaia* bands there is even a strong dependence on the temperature for the Cepheids (and probably also on the reddening) for a given reddening law.

**Table 9.** Relative reddening values for different reddening laws for Models 1/2/3.

| Filter | $\lambda_{\text{eff}}$/Model | F99 | COD | W23 |
|--------|------|-----|-----|-----|
| | | $R_V = 3.3$ | $R_V = 3.1$ | $R_V = 3.4$ |
| | | | $A_\lambda/A_V$ | |
| $G$ | 7020/7690/6580 | 0.697/0.601/0.769 | 0.751/0.655/0.810 | 0.703/0.610/0.776 |
| $Bp$ | 5630/5780/5380 | 0.950/0.919/1.008 | 0.969/0.937/1.029 | 0.968/0.935/1.027 |
| $Rp$ | 7940/8260/7850 | 0.570/0.532/0.581 | 0.617/0.569/0.631 | 0.579/0.543/0.590 |
| $V_{\text{o}}$ | 5410/5430/5370 | 1.001/0.996/1.011 | 1.021/1.016/1.032 | 1.020/1.015/1.030 |
| $I_{\text{o}}$ | 7980/8030/7960 | 0.565/0.559/0.567 | 0.611/0.603/0.614 | 0.575/0.569/0.577 |
| $J$ | 12370/12400/12310 | 0.261/0.260/0.263 | 0.287/0.286/0.289 | 0.249/0.247/0.252 |
| $H$ | 16430/16480/16410 | 0.165/0.164/0.166 | 0.182/0.181/0.182 | 0.138/0.138/0.139 |
| $K_{\text{s}}$ | 21530/21540/21530 | 0.112/0.112/0.112 | 0.118/0.117/0.118 | 0.079/0.079/0.079 |
| F555W | 5450/5480/5400 | 1.011/0.984/1.003 | 1.011/1.004/1.024 | 1.011/1.003/1.022 |
| F814W | 8140/8280/8100 | 0.587/0.530/0.551 | 0.587/0.566/0.593 | 0.557/0.541/0.561 |
| F160W | 15320/15410/15270 | 0.203/0.182/0.185 | 0.203/0.201/0.204 | 0.160/0.158/0.161 |
| | | | $R$ | |
| $W_G$ | | 1.831/1.557/1.798 | 2.136/1.781/2.034 | 1.807/1.558/1.775 |
| $W_{VI}$ | | 1.296/1.279/1.279 | 1.490/1.463/1.471 | 1.291/1.275/1.276 |
| $W_{JK}$ | | 0.751/0.756/0.740 | 0.694/0.698/0.685 | 0.465/0.468/0.459 |
| $W_H$ | | 0.414/0.402/0.409 | 0.479/0.461/0.474 | 0.352/0.342/0.349 |

What is the effect in practice, and how does one deal with Wesenheit coefficients that differ per star? One can still take a fiducial relation, say a Wesenheit relation and consider it to be just another period–luminosity–colour relation where you need to deredden explicitly, and calibrate against that. Table 10 gives an example for CR Car which is a 9.7 day DCEP. The table lists the observed magnitudes (the reddening coefficients for the OGLE filters were used for $V$ and $I$), the dereddened magnitudes, and the unreddened and dereddened magnitudes for the 'period–luminosity–colour' relations, $W_G = G - 1.90(Bp - Rp)$, $W_{VI} = I - 1.387(V - I)$, $W_{JK} = K - 0.735(J - K)$, and $W_H = \text{F160W} - 0.386(\text{F555W} - \text{F814W})$.

In this particular case, assuming that the adopted Wesenheit relation takes care of reddening and is universal and thus using the observed magnitudes, compared to actually dereddening the photometry assuming different reddening values introduces differences that can be as large as almost 0.1 mag, especially in the optical. In the IR the effect is smaller but may still be 0.05 mag. The exact effects will depend on a star-to-star basis and the choice of the fiducial period–luminosity–colour relations.

The effect of different reddening laws will also potentially affect other standard candles, e.g., the tip of the red-giant branch method that requires $I$ (and a $(V - I)$ colour), or the JAGB method that requires $J$ (and a $(J - K)$ colour). For example, the often used reddening map of the Magellanic Clouds of Skowron et al. (2021) is given in terms of $E(V - I)$ and needs to be transformed to the reddening in other bands†

## 3. Galactic structure

Once accurate and precise PL relations are available, distances can be computed (and those will be more accurate than distances derived from the currently available *Gaia* data, particularly at larger distances) and an obvious application is in the context of Galactic structure. Many other tracers than DCEPs and RR Lyrae have been used to study Galactic structure (e.g., OB stars, OCs) but those will not be covered here. More extended reviews on this topic were given by Dorota Skowron and Vasiliy Belokurov at this conference.

---

† The value to transform $E(V - I)$ to $E(B - V)$ given by Skowron et al. (2021) is only valid for one particular reddening law.

**Table 10.** Observed and dereddened magnitudes for CR Car and different reddening laws.

| Filter | Observed | Dereddened magnitudes | | |
|---|---|---|---|---|
| | | F99/ Model 1 | COD/ Model 1 | W23/ Model 3 |
| | | $R_V = 3.3$ | $R_V = 3.1$ | $R = 3.4$ |
| | | $E(B-V) = 0.5$ | $E(B-V) = 0.5$ | $E(B-V) = 0.1$ |
| $G$ | 11.063 | 9.909 | 9.875 | 10.655 |
| $Bp$ | 11.880 | 10.311 | 10.347 | 11.341 |
| $Rp$ | 10.137 | 9.199 | 9.161 | 9.827 |
| $V$ | 11.511 | 9.859 | 9.928 | 10.986 |
| $I$ | 10.007 | 9.074 | 9.040 | 9.704 |
| $J$ | 8.846 | 8.415 | 8.392 | 8.714 |
| $H$ | 8.188 | 7.916 | 7.900 | 8.115 |
| $K$ | 8.058 | 7.873 | 7.871 | 8.017 |
| F555W | 11.750 | 10.080 | 10.150 | 11.213 |
| F814W | 9.973 | 9.003 | 9.044 | 9.678 |
| F160W | 8.384 | 8.049 | 8.063 | 8.299 |
| $W_G$ | 7.751 | 7.795 | 7.621 | 7.780 |
| $W_{VI}$ | 7.921 | 7.984 | 7.808 | 7.926 |
| $W_{JK}$ | 7.479 | 7.475 | 7.489 | 7.504 |
| $W_H$ | 7.698 | 7.633 | 7.636 | 7.707 |

### 3.1. *Distance to the Galactic Centre*

An obvious application is to use variable stars to determine the structure of the Galactic Centre (GC) region and in particular the distance to the GC. Majaess et al. (2018) used 4194 RRab stars from the VISTA Variables in the Vía Láctea survey (VVV; Minniti et al. 2010), in the region $350 \lesssim l \lesssim 10°$, $-10 \lesssim b \lesssim$ -2.7°, $2.7 \lesssim b \lesssim 5°$, an $M_K$ (based on an LMC distance modulus of 18.43 mag and no metallicity term) and a $(J - K)_0$ versus $\log P$ relation to derive $d_{GC} = 8.30 \pm 0.36$ kpc (for a subsample with $|b| > 4°$), uncorrected for the cone effect.

Braga et al. (2018) presented IR data for 894 T2C in the VVV survey (with periods from OGLE) and derived an $M_K - \log P$ relation that also includes terms depending on Galactic longitude and latitude. They obtained $d_{GC} = 8.46 \pm 0.14$ kpc (combining statistical and systematic errors). Griv et al. (2021) used 715 high-latitude ($|b| > 1°$) and centrally symmetrically concentrated T2Cs identified in the VVV survey to derive $d_{GC} = 8.35 \pm 0.10$ kpc, concluding that the T2C distribution is an ellipsoid with axial ratios of about 1:0.7:0.6.

These distances agree with that derived from the stars orbiting Sgr A*, $d_{GC} = 8277 \pm 34$ pc (adding random and systematic errors; see GRAVITY Collaboration et al. 2022, 2021, 2019).

### 3.2. *RR Lyrae*

Pietrukowicz et al. (2020) used the OGLE survey to discriminate between MW and background RRL to study samples of 53,000 Galactic bulge and disc and 951 halo RRab stars (plus 2288 Sgr dSph, 26,900 LMC and 4600 SMC RRL). They used photometric metallicities to derive characteristic metallicites for the halo/bulge/disc of [Fe/H] = $-1.2, -1.0$, and $-0.6$ dex, respectively. They found that 1/3 of the RRL within the bulge area actually belong to the halo.

Iorio and Belokurov (2021) used 5D *Gaia* DR2 data to study the kinematical properties of halo and disc RRL to find a dominant, non-rotating, halo-like population and a much smaller rotating disc-like population.

Li et al. (2023) used a sample of 2700 RRL with precise metallicities to calibrate a $P$–$\phi_{31}$–$R_{21}$–[Fe/H] relation, and 205 RRab and 31 RRc to calibrate an $M_G$–[Fe/H] relation using parallaxes from GDR3. This was then applied to 115,410 RRab and 20,463 RRc stars in GDR3 and used to derive the distances to the LMC and SMC, the mean metallicities and metallicity gradients in the Magellanic Clouds, and the values were compared with independent metallicity and distance determinations of RRL in globular clusters.

### 3.3. *Classical Cepheids*

Classical Cepheids, being quite young objects, can be used to study the properties of the spiral arms in the MW, the presence and properties of the warp, when velocities are available, properties of the Galactic rotation curve, and the metallicity gradient in the Galactic disc. An updated list of Galactic Cepheids is given by Pietrukowicz et al. (2021)†.

Minniti et al. (2021) use 50 classical Cepheids from the VVV to study the spiral structure in the poorly explored far side of the disc (quadrants I and IV).

Ablimit et al. (2020) use classical Cepheids from several photometric surveys (OGLE, ASAS-SN, *Gaia*, *WISE*, and the Zwicky Transient Facility) and calculate distances based on the *WISE* PL relations from Wang et al. (2018); 3483 objects with distances from the Galactic plane of less than 4 kpc are kept. The warp is clearly visible (as already seen in earlier work). Using PMs from GDR2 and RVs from GDR2 and LAMOST DR6 applied to a cleaned sample of 968 DCEPs they derive a rotation curve of the form $\Theta(R) = (232.5 \pm 0.8) + (-1.3 \pm 0.1)(R - R_0)$ km s$^{-1}$ (for $R_0 = 8.122$ kpc). Kinematics studies have also been performed by Bobylev and Bajkova (2023) (also by Bobylev et al. 2021) but using fewer (2000) stars and PL relations that ultimately are tied GDR2. Similar studies, not only using DCEPs but also OB stars, OCs, and RGB stars, are Gaia Collaboration et al. (2023a) and Drimmel et al. (2023).

Lemasle et al. (2022) studied the warp and spiral arms using unWISE (Meisner et al. 2021) photometry for 3260 MW classical Cepheids and a Wesenheit index using *WISE* filters, $W_{W1W2} = W2 - 2.0(W1 - W2)$, calibrated on a sample of LMC DCEPs and the canonical LMC distance. This was applied to the MW sample without any correction for metallicity. They then fitted a model of the form (Skowron et al. 2019)

$$z(r, \Theta) = \begin{cases} z_0 & R_{GC} < r_0 \\ z_0 + (R_{GC} - r_0)^2 \cdot [z_1 \sin(\Theta - \Theta_1) + z_2 \sin(2(\Theta - \Theta_2))] & R_{GC} \geq r_0, \end{cases}$$

where $z$ is the vertical distance from the Galactic plane, $R_{GC}$ is the distance from the GC, and $\Theta$ the Galactocentric azimuth. Warping starts at $R_{GC} \sim 10$–12 kpc. Earlier studies have been carried out by Chen et al. (2019) and Skowron et al. (2019), while another very recent study is Dehnen et al. (2023), who also conclude that there is no warping inside about 11 kpc.

Another classical application is to study abundance patterns throughout the MW. Trentin et al. (2023) use 637 DCEPs and a $M_{W_G}$ relation to derive [Fe/H] $= (-0.060 \pm 0.002)R_{GC} + (0.573 \pm 0.017)$ dex (for $R_{GC} > 4$ kpc), with a possible break at 9.25 kpc.

Hocdé et al. (2023) took the interesting approach to derive [Fe/H] from the Cepheids' light curve Fourier decomposition (a standard method for RRL). For short-period Cepheids ($2.5 < P < 6.3$ days) they suggest to use a relation based on the amplitudes $A_1$ and $A_2$, and for long-period Cepheids ($12 < P < 40$ days) a relation using $A_1$, $\phi_{21}$, and $R_{41}$. The metallicity gradient they derive is in agreement with the literature. Lemasle et al. (2022) show the radial gradient as well and refer to the companion paper by da Silva et al. (2022) who give [Fe/H] $= (-0.055 \pm$ $0.003)R_{GC} + (+0.43 \pm 0.03)$. Kovtyukh et al. (2022) also present the metallicity gradient, but they do not cite a linear coefficient and their figure 2 seems to suggest a slope that steepens in the inner disc.

---

† 3666 as per the version of September 2022; see `https://www.astrouw.edu.pl/ogle/ogle4/OCVS/allGalCep.listID`

## 4. Conclusions

Compared to a few years ago, the metallicity dependence of the Cepheid PL relations is now better determined. What remains to be seen if the dependence derived using the SMC, LMC, and using a single average metallicity for the MW, and that derived from MW DCEPs using the individually derived metallicities can be brought into better agreement.

Ongoing and future spectroscopic surveys in both the optical and the near-IR (NIR; ground-based as well as future *Gaia* releases) will provide many more accurate metallicity determinations (as well as radial velocities) that will help in addressing this question, as well as be useful to study abundance gradients (radial and vertical) in more detail and for kinematics studies. Reddening remains an important component, and blindly using Wesenheit indices may lead to systematic effects.

Two topics have not been addressed here. The first is the role of binarity, leading to observed magnitudes that are those of the Cepheid plus a companion. In terms of calibration it becomes a differential effect. This was recently tackled using population synthesis (Karczmarek et al. 2022, 2023) with the conclusion that there is essentially no effect on the derived slope of the PL relation, and a small effect on the zeropoint, depending on filter and binary fraction. However, further investigations are required.

The second point is the role of any excess emission in the NIR and mid-IR. This is known to exist in Galactic DCEPs, e.g., from direct interferometric observations in the optical or NIR (e.g., Kervella et al. 2006; Mérand et al. 2006; Gallenne et al. 2012; Nardetto et al. 2016), modelling with the SPIPS code (e.g., Breitfelder et al. 2016; Trahin 2019; Trahin et al. 2021), or modelling the SEDs of Galactic DCEPs (Gallenne et al. 2013; Groenewegen 2020). Groenewegen (2020) found evidence of an IR excess in 5% of the sample. Groenewegen & Lub (in prep.) studied the spectral energy distributions of LMC and SMC DCEPs and found evidence of an IR excess in 1/142 LMC and 0/77 SMC Cepheids studied. However, there is less mid- and far-IR data available for the Magellanic Clouds, and so it is harder to detect significant excess emission than for MW DCEPs. At face value, a lower percentage of stars with IR excess is consistent with the fact that there is less condensible material available in the LMC and SMC. However, the interpretation of the IR excess in terms of dust emission has also been questioned, and Hocdé et al. (2020) proposed an alternative origin, namely free–free emission from a thin hydrogen shell. In that case, the abundance argument would not play an important role.

A final comment relates to the 'countercorrections' (Eq. 2) that have been derived in the recent literature, i.e., the additional correction one has to apply after applying the L21 correction to bring the parallaxes of one's favourite sample of objects in agreement with an absolute calibration relation for that specific class. Figure 3 collects recent determinations related to GDR3. Over a large range in $G$ magnitude the derived countercorrections do not agree. Khan et al. (2023) also showed that $\Delta\pi$ depends on sky position.

Uncorrected parallaxes have offsets of $-17$ $\mu$as (Lindegren et al. 2021a) at the faint magnitudes of the QSOs, or $\sim -24$ to $-28$ $\mu$as (Riess et al. 2021b; Huang et al. 2021b; Ren et al. 2021; Wang et al. 2022) at brighter magnitudes. This implies that some countercorrections almost completely counter the L21 correction itself, which is hard to understand given its physical basis.

**Acknowledgements.** I thank Mauricio Cruz Reyes for kindly making the membership lists of OCs containing Cepheids available before publication.

This work has made use of data from the European Space Agency (ESA) mission *Gaia* (`https://www.cosmos.esa.int/gaia`), processed by the *Gaia* Data Processing and Analysis Consortium (DPAC, `https://www.cosmos.esa.`

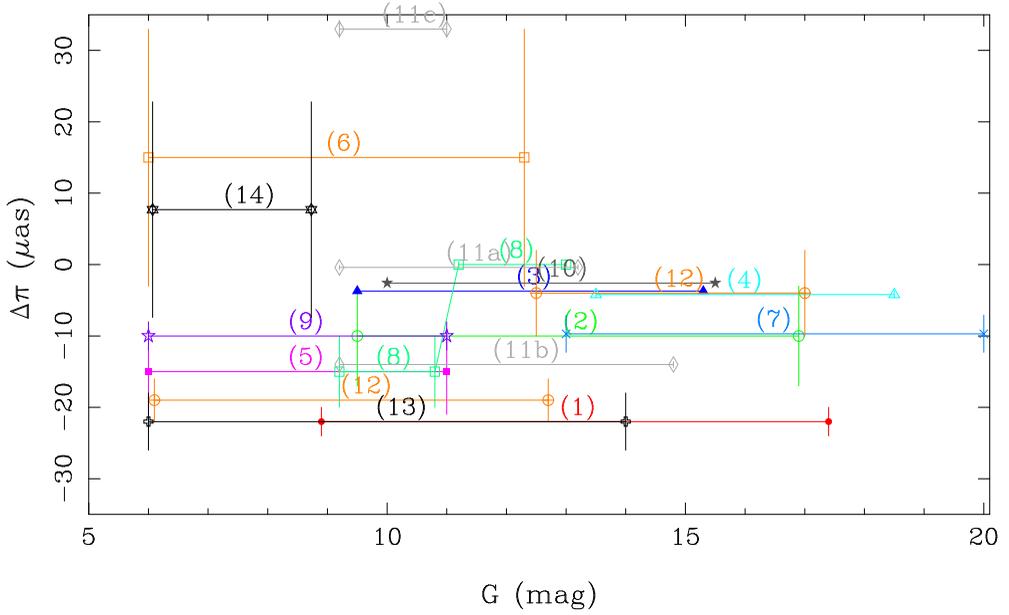

**Figure 3.** Countercorrection $\Delta\pi$ plotted against magnitude for (1) Bhardwaj et al. (2021), RRL; (2) Gilligan et al. (2021), RRL; (3) Huang et al. (2021b), red clump (5p solution); (4) Ren et al. (2021), WUMa EBs (5p solution); (5) Riess et al. (2021a), DCEP LMC; (6) Stassun and Torres (2021), EBs; (7) Vasiliev and Baumgardt (2021), globular clusters; (8) Zinn (2021), asteroseismology; (9) Flynn et al. (2022), open/globular clusters; $Bp - Rp > 1$ (10) Wang et al. (2022), giants; (11) Khan et al. (2023) red giants: asteroseismology, for the (a) *Kepler*, (b) *K2*, and (c) *TESS*-SCVZ fields ('E20', APOGEE DR17 values); (12) Cruz Reyes and Anderson (2023) Clusters ($G = 12.5 - 17$ mag, $0.8 < Bp - Rp < 2.75$ mag) and MW DCEPs; (13) Molinaro et al. (2023) MW DCEPS; (14) Groenewegen (2023) dynamical parallax of binary systems with spectroscopic and astrometric orbits.

int/web/gaia/dpac/consortium). Funding for the DPAC has been provided by national institutions, in particular the institutions participating in the *Gaia* Multilateral Agreement.

## References

Ablimit, I., Zhao, G., Flynn, C., & Bird, S. A. 2020, *ApJL*, 895, L12.

Anderson, R. I., Eyer, L., & Mowlavi, N. 2013, *MNRAS*, 434, 2238–2261.

Bailer-Jones, C. A. L., Rybizki, J., Fouesneau, M., Demleitner, M., & Andrae, R. 2021, *AJ*, 161, 147.

Baumgardt, H. & Vasiliev, E. 2021, *MNRAS*, 505, 5957–5977.

Benedict, G. F., Barnes, T. G., Evans, N. R., Cochran, W. D., Anderson, R. I., McArthur, B. E., & Harrison, T. E. 2022, *AJ*, 163, 282.

Benedict, G. F., McArthur, B. E., Feast, M. W., Barnes, T. G., Harrison, T. E., Bean, J. L., Menzies, J. W., Chaboyer, B., Fossati, L., Nesvacil, N., Smith, H. A., Kolenberg, K., Laney, C. D., Kochukhov, O., Nelan, E. P., Shulyak, D. V., Taylor, D., & Freedman, W. L. 2011, *AJ*, 142, 187.

Benedict, G. F., McArthur, B. E., Feast, M. W., Barnes, T. G., Harrison, T. E., Patterson, R. J., Menzies, J. W., Bean, J. L., & Freedman, W. L. 2007, *AJ*, 133, 1810–1827.

Benedict, G. F., McArthur, B. E., Fredrick, L. W., Harrison, T. E., Lee, J., Slesnick, C. L., Rhee, J., Patterson, R. J., Nelan, E., Jefferys, W. H., van Altena, W., Shelus, P. J., Franz, O. G., Wasserman, L. H., Hemenway, P. D., Duncombe, R. L., Story, D., Whipple, A. L., & Bradley, A. J. 2002, *AJ*, 123, 473–484.

Benedict, G. F., McArthur, B. E., Nelan, E. P., & Harrison, T. E. 2017, *PASP*, 129, 012001.

Bhardwaj, A., Marconi, M., Rejkuba, M., de Grijs, R., Singh, H. P., Braga, V. F., Kanbur, S., Ngeow, C.-C., Ripepi, V., Bono, G., De Somma, G., & Dall'Ora, M. 2023, *ApJL*, 944, L51.

Bhardwaj, A., Rejkuba, M., de Grijs, R., Yang, S.-C., Herczeg, G. J., Marconi, M., Singh, H. P., Kanbur, S.,

& Ngeow, C.-C. 2021, *ApJ*, 909, 200.

Bobylev, V. V. & Bajkova, A. T. 2023, *RAA*, 23, 045001.

Bobylev, V. V., Bajkova, A. T., Rastorguev, A. S., & Zabolotskikh, M. V. 2021, *MNRAS*, 502, 4377.

Braga, V. F., Bhardwaj, A., Contreras Ramos, R., Minniti, D., Bono, G., de Grijs, R., Minniti, J. H., & Rejkuba, M. 2018, *A&A*, 619, A51.

Breitfelder, J., Mérand, A., Kervella, P., Gallenne, A., Szabados, L., Anderson, R. I., & Le Bouquin, J.-B. 2016, *A&A*, 587, A117.

Breuval, L., Kervella, P., Anderson, R. I., Riess, A. G., Arenou, F., Trahin, B., Mérand, A., Gallenne, A., Gieren, W., Storm, J., Bono, G., Pietrzyński, G., Nardetto, N., Javanmardi, B., & Hocdé, V. 2020, *A&A*, 643, A115.

Breuval, L., Riess, A. G., Kervella, P., Anderson, R. I., & Romaniello, M. 2022, *ApJ*, 939, 89.

Cantat-Gaudin, T., Jordi, C., Vallenari, A., Bragaglia, A., Balaguer-Núñez, L., Soubiran, C., Bossini, D., Moitinho, A., Castro-Ginard, A., Krone-Martins, A., Casamiquela, L., Sordo, R., & Carrera, R. 2018, *A&A*, 618, A93.

Cardelli, J. A., Clayton, G. C., & Mathis, J. S. 1989, *ApJ*, 345, 245.

Chen, X., de Grijs, R., & Deng, L. 2017, *MNRAS*, 464, 1119.

Chen, X., Wang, S., Deng, L., de Grijs, R., Liu, C., & Tian, H. 2019, *Nat. Astron.*, 3, 320.

Cruz Reyes, M. & Anderson, R. I. 2023, *A&A*, 672, A85.

da Silva, R., Crestani, J., Bono, G., Braga, V. F., D'Orazi, V., Lemasle, B., Bergemann, M., Dall'Ora, M., Fiorentino, G., François, P., Groenewegen, M. A. T., Inno, L., Kovtyukh, V., Kudritzki, R. P., Matsunaga, N., Monelli, M., Pietrinferni, A., Porcelli, L., Storm, J., Tantalo, M., & Thévénin, F. 2022, *A&A*, 661, A104.

Dehnen, W., Semczuk, M., & Schönrich, R. 2023, *MNRAS*, 523, 1556–1564.

Drimmel, R., Khanna, S., D'Onghia, E., Tepper-García, T., Bland-Hawthorn, J., Chemin, L., Ripepi, V., Romero-Gómez, M., Ramos, P., Poggio, E., Andrae, R., Blomme, R., Cantat-Gaudin, T., Castro-Ginard, A., Clementini, G., Figueras, F., Fouesneau, M., Frémat, Y., Lobel, A., Marshall, D., & Muraveva, T. 2023, *A&A*, 670, A10.

ESA 1997, *VizieR Online Data Catalog*, 1239.

Fabricius, C., Luri, X., Arenou, F., Babusiaux, C., Helmi, A., Muraveva, T., Reylé, C., Spoto, F., Vallenari, A., Antoja, T., Balbinot, E., Barache, C., Bauchet, N., Bragaglia, A., Busonero, D., Cantat-Gaudin, T., Carrasco, J. M., Diakité, S., Fabrizio, M., Figueras, F., Garcia-Gutierrez, A., Garofalo, A., Jordi, C., Kervella, P., & et al. 2021, *A&A*, 649, A5.

Feast, M. W. & Catchpole, R. M. 1997, *MNRAS*, 286, L1.

Feast, M. W., Laney, C. D., Kinman, T. D., van Leeuwen, F., & Whitelock, P. A. 2008, *MNRAS*, 386, 2115.

Fitzpatrick, E. L. 1999, *PASP*, 111, 63.

Flynn, C., Sekhri, R., Venville, T., Dixon, M., Duffy, A., Mould, J., & Taylor, E. N. 2022, *MNRAS*, 509, 4276–4284.

Fouqué, P., Arriagada, P., Storm, J., Barnes, T. G., Nardetto, N., Mérand, A., Kervella, P., Gieren, W., Bersier, D., Benedict, G. F., & McArthur, B. 2007, *A&A*, 476, 73.

Gaia Collaboration, Brown, A. G. A., Vallenari, A., Prusti, T., de Bruijne, J. H. J., Babusiaux, C., Biermann, M., Creevey, O. L., Evans, D. W., Eyer, L., Hutton, A., Jansen, F., Jordi, C., Klioner, S. A., Lammers, U., Lindegren, L., Luri, X., Mignard, F., Panem, C., Pourbaix, D., Randich, S., & et al. 2021, *A&A*, 649, A1.

Gaia Collaboration, Brown, A. G. A., Vallenari, A., Prusti, T., de Bruijne, J. H. J., Mignard, F., Drimmel, R., Babusiaux, C., Bailer-Jones, C. A. L., Bastian, U., & et al. 2016,a *A&A*, 595a, A2.

Gaia Collaboration, Drimmel, R., Romero-Gomez, M., Chemin, L., Ramos, P., Poggio, E., Ripepi, V., Andrae, R., Blomme, R., Cantat-Gaudin, T., Castro-Ginard, A., Clementini, G., Figueras, F., Fouesneau, M., Fremat, Y., Jardine, K., Khanna, S., Lobel, A., Marshall, D. J., Muraveva, T., Brown, A. G. A., Vallenari, A., Prusti, T., de Bruijne, J. H. J., Arenou, F., Babusiaux, C., Biermann, M., & et al. 2023,a *A&A*, 674a, A37.

Gaia Collaboration, Prusti, T., de Bruijne, J. H. J., Brown, A. G. A., Vallenari, A., Babusiaux, C., Bailer-Jones, C. A. L., Bastian, U., Biermann, M., Evans, D. W., & et al. 2016,b *A&A*, 595b, A1.

Gaia Collaboration, Vallenari, A., Brown, A.G.A., Prusti, T., & et al. 2023,b *A&A*, 674b, A1.

Gallenne, A., Kervella, P., & Mérand, A. 2012, *A&A*, 538, A24.

Gallenne, A., Mérand, A., Kervella, P., Chesneau, O., Breitfelder, J., & Gieren, W. 2013, *A&A*, 558, A140.

Garofalo, A., Delgado, H. E., Sarro, L. M., Clementini, G., Muraveva, T., Marconi, M., & Ripepi, V. 2022, *MNRAS*, 513, 788.

Gieren, W., Storm, J., Konorski, P., Górski, M., Pilecki, B., Thompson, I., Pietrzyński, G., Graczyk, D., Barnes, T. G., Fouqué, P., Nardetto, N., Gallenne, A., Karczmarek, P., Suchomska, K., Wielgórski, P., Taormina, M., & Zgirski, B. 2018, *A&A*, 620, A99.

Gilligan, C. K., Chaboyer, B., Marengo, M., Mullen, J. P., Bono, G., Braga, V. F., Crestani, J., Dall'Ora, M., Fiorentino, G., Monelli, M., Neeley, J. R., Fabrizio, M., Martínez-Vázquez, C. E., Thévenin, F., & Sneden, C. 2021, *MNRAS*, 503, 4719–4733.

Górski, K. M., Hivon, E., Banday, A. J., Wandelt, B. D., Hansen, F. K., Reinecke, M., & Bartelmann, M. 2005, *ApJ*, 622, 759.

Gratton, R. G. 1998, *MNRAS*, 296, 739.

GRAVITY Collaboration, Abuter, R., Aimar, N., Amorim, A., Ball, J., Bauböck, M., Berger, J. P., Bonnet, H., Bourdarot, G., Brandner, W., Cardoso, V., Clénet, Y., Dallilar, Y., Davies, R., de Zeeuw, P. T., Dexter, J., Drescher, A., Eisenhauer, F., Förster Schreiber, N. M., Foschi, A., Garcia, P., Gao, F., Gendron, E., Genzel, R., Gillessen, S., Habibi, M., Haubois, X., Heißel, G., Henning, T., Hippler, S., Horrobin, M., Jochum, L., Jocou, L., Kaufer, A., Kervella, P., Lacour, S., Lapeyrère, V., Le Bouquin, J. B., Léna, P., Lutz, D., Ott, T., Paumard, T., Perraut, K., Perrin, G., Pfuhl, O., Rabien, S., Shangguan, J., Shimizu, T., Scheithauer, S., Stadler, J., Stephens, A. W., Straub, O., Straubmeier, C., Sturm, E., Tacconi, L. J., Tristram, K. R. W., Vincent, F., von Fellenberg, S., Widmann, F., Wieprecht, E., Wiezorrek, E., Woillez, J., Yazici, S., & Young, A. 2022, *A&A*, 657, L12.

GRAVITY Collaboration, Abuter, R., Amorim, A., Bauböck, M., Berger, J. P., Bonnet, H., Brandner, W., Clénet, Y., Coudé Du Foresto, V., de Zeeuw, P. T., Dexter, J., Duvert, G., Eckart, A., Eisenhauer, F., Förster Schreiber, N. M., Garcia, P., Gao, F., Gendron, E., Genzel, R., Gerhard, O., Gillessen, S., Habibi, M., Haubois, X., Henning, T., Hippler, S., Horrobin, M., Jiménez-Rosales, A., Jocou, L., Kervella, P., & et al. 2019, *A&A*, 625, L10.

GRAVITY Collaboration, Abuter, R., Amorim, A., Bauböck, M., Berger, J. P., Bonnet, H., Brandner, W., Clénet, Y., Davies, R., de Zeeuw, P. T., Dexter, J., Dallilar, Y., Drescher, A., Eckart, A., Eisenhauer, F., Förster Schreiber, N. M., Garcia, P., Gao, F., Gendron, E., Genzel, R., Gillessen, S., Habibi, M., Haubois, X., Heißel, G., Henning, T., Hippler, S., Horrobin, M., Jiménez-Rosales, A., Jochum, L., Jocou, L., & et al. 2021, *A&A*, 647, A59.

Griv, E., Gedalin, M., Pietrukowicz, P., Majaess, D., & Jiang, I.-G. 2021, *MNRAS*, 502, 4194.

Groenewegen, M. A. T. In Recio-Blanco, A., de Laverny, P., Brown, A. G. A., & Prusti, T., editors, *Astrometry and Astrophysics in the Gaia Sky* 2018,a, volume 330, 287.

Groenewegen, M. A. T. 2018,b *A&A*, 619b, A8.

Groenewegen, M. A. T. 2020, *A&A*, 635, A33.

Groenewegen, M. A. T. 2021, *A&A*, 654, A20.

Groenewegen, M. A. T. 2023, *A&A*, 669, A4.

Gustafsson, B., Edvardsson, B., Eriksson, K., Jørgensen, U. G., Nordlund, Å., & Plez, B. 2008, *A&A*, 486, 951.

Hao, C. J., Xu, Y., Wu, Z. Y., Lin, Z. H., Bian, S. B., Li, Y. J., & Liu, D. J. 2022, *A&A*, 668, A13.

Hocdé, V., Nardetto, N., Lagadec, E., Niccolini, G., Domiciano de Souza, A., Mérand, A., Kervella, P., Gallenne, A., Marengo, M., Trahin, B., Gieren, W., Pietrzyński, G., Borgniet, S., Breuval, L., & Javanmardi, B. 2020, *A&A*, 633, A47.

Hocdé, V., Smolec, R., Moskalik, P., Ziółkowska, O., & Singh Rathour, R. 2023, *A&A*, 671, A157.

Huang, Y., Li, Q., Zhang, H., Li, X., Sun, W., Chang, J., Dong, X., & Liu, X. 2021,a *ApJL*, 907a, L42.

Huang, Y., Yuan, H., Beers, T. C., & Zhang, H. 2021,b *ApJL*, 910b, L5.

Hunt, E. L. & Reffert, S. 2023, *A&A*, 673, A114.

Iorio, G. & Belokurov, V. 2021, *MNRAS*, 502, 5686–5710.

Karczmarek, P., Hajdu, G., Pietrzyński, G., Gieren, W., Narloch, W., Smolec, R., Wiktorowicz, G., & Bełczynski, K. 2023, *arXiv e-prints*, arXiv:2303.15664.

Karczmarek, P., Smolec, R., Hajdu, G., Pietrzyński, G., Gieren, W., Narloch, W., Wiktorowicz, G., & Bełczynski, K. 2022, *ApJ*, 930, 65.

Kervella, P., Bond, H. E., Cracraft, M., Szabados, L., Breitfelder, J., Mérand, A., Sparks, W. B., Gallenne, A., Bersier, D., Fouqué, P., & Anderson, R. I. 2014, *A&A*, 572, A7.

Kervella, P., Mérand, A., Perrin, G., & Coudé du Foresto, V. 2006, *A&A*, 448, 623.

Khan, S., Miglio, A., Willett, E., Mosser, B., Elsworth, Y. P., Anderson, R. I., Girardi, L., Belkacem, K., Brown, A. G. A., Cantat-Gaudin, T., Casagrande, L., Clementini, G., & Vallenari, A. 2023, *arXiv e-prints*,, arXiv:2304.07158.

Kovtyukh, V., Lemasle, B., Bono, G., Usenko, I. A., da Silva, R., Kniazev, A., Grebel, E. K., Andronov, I. L., Shakun, L., & Chinarova, L. 2022, *MNRAS*, 510, 1894.

Layden, A. C., Tiede, G. P., Chaboyer, B., Bunner, C., & Smitka, M. T. 2019, *AJ*, 158, 105.

Lemasle, B., Lala, H. N., Kovtyukh, V., Hanke, M., Prudil, Z., Bono, G., Braga, V. F., da Silva, R., Fabrizio, M., Fiorentino, G., François, P., Grebel, E. K., & Kniazev, A. 2022, *A&A*, 668, A40.

Li, X.-Y., Huang, Y., Liu, G.-C., Beers, T. C., & Zhang, H.-W. 2023, *ApJ*, 944, 88.

Lin, Z., Xu, Y., Hao, C., Liu, D., Li, Y., & Bian, S. 2022, *ApJ*, 938, 33.

Lindegren, L., Bastian, U., Biermann, M., Bombrun, A., de Torres, A., Gerlach, E., Geyer, R., Hernández, J., Hilger, T., Hobbs, D., Klioner, S. A., Lammers, U., McMillan, P. J., Ramos-Lerate, M., Steidelmüller, H., Stephenson, C. A., & van Leeuwen, F. 2021,a *A&A*, 649a, A4.

Lindegren, L., Hernández, J., Bombrun, A., Klioner, S., Bastian, U., Ramos-Lerate, M., de Torres, A., Steidelmüller, H., Stephenson, C., Hobbs, D., Lammers, U., Biermann, M., Geyer, R., Hilger, T., Michalik, D., Stampa, U., McMillan, P. J., Castañeda, J., Clotet, M., Comoretto, G., Davidson, M., Fabricius, C., Gracia, G., Hambly, N. C., Hutton, A., Mora, A., Portell, J., van Leeuwen, F., & et al. 2018, *A&A*, 616, A2.

Lindegren, L., Klioner, S. A., Hernández, J., Bombrun, A., Ramos-Lerate, M., Steidelmüller, H., Bastian, U., Biermann, M., de Torres, A., Gerlach, E., Geyer, R., Hilger, T., Hobbs, D., Lammers, U., McMillan, P. J., Stephenson, C. A., Castañeda, J., Davidson, M., Fabricius, C., Gracia-Abril, G., Portell, J., Rowell, N., Teyssier, D., Torra, F., Bartolomé, S., Clotet, M., Garralda, N., González-Vidal, J. J., Torra, J., & et al. 2021,b *A&A*, 649b, A2.

Looijmans, K., Lub, J., & Brown, A. G. A. 2023, *arXiv e-prints*,, arXiv:2303.02211.

Madore, B. F. 1982, *ApJ*, 253, 575.

Madore, B. F. & Freedman, W. L. 1998, *ApJ*, 492, 110.

Maíz Apellániz, J. 2022, *A&A*, 657, A130.

Maíz Apellániz, J., Pantaleoni González, M., & Barbá, R. H. 2021,.

Majaess, D., Dékány, I., Hajdu, G., Minniti, D., Turner, D., & Gieren, W. 2018, *Ap&SS*, 363, 127.

Medina, G. E., Lemasle, B., & Grebel, E. K. 2021, *MNRAS*, 505, 1342.

Meisner, A. M., Lang, D., Schlafly, E. F., & Schlegel, D. J. 2021, *RNAAS*, 5, 168.

Mérand, A., Kervella, P., Coudé du Foresto, V., Perrin, G., Ridgway, S. T., Aufdenberg, J. P., ten Brummelaar, T. A., McAlister, H. A., Sturmann, L., Sturmann, J., Turner, N. H., & Berger, D. H. 2006, *A&A*, 453, 155.

Minniti, D., Lucas, P. W., Emerson, J. P., Saito, R. K., Hempel, M., Pietrukowicz, P., Ahumada, A. V., Alonso, M. V., Alonso-Garcia, J., Arias, J. I., Bandyopadhyay, R. M., Barbá, R. H., Barbuy, B., Bedin, L. R., Bica, E., Borissova, J., Bronfman, L., Carraro, G., Catelan, M., Clariá, J. J., Cross, N., de Grijs, R., Dékány, I., Drew, J. E., Fariña, C., Feinstein, C., Fernández Lajús, E., Gamen, R. C., Geisler, D., Gieren, W., Goldman, B., Gonzalez, O. A., Gunthardt, G., Gurovich, S., Hambly, N. C., Irwin, M. J., Ivanov, V. D., Jordán, A., Kerins, E., Kinemuchi, K., Kurtev, R., López-Corredoira, M., Maccarone, T., Masetti, N., Merlo, D., Messineo, M., Mirabel, I. F., Monaco, L., Morelli, L., Padilla, N., Palma, T., Parisi, M. C., Pignata, G., Rejkuba, M., Roman-Lopes, A., Sale, S. E., Schreiber, M. R., Schröder, A. C., Smith, M., , Jr., L. S., Soto, M., Tamura, M., Tappert, C., Thompson, M. A., Toledo, I., Zoccali, M., & Pietrzynski, G. 2010, *NewA*, 15, 433.

Minniti, J. H., Zoccali, M., Rojas-Arriagada, A., Minniti, D., Sbordone, L., Contreras Ramos, R., Braga, V. F., Catelan, M., Duffau, S., Gieren, W., Marconi, M., & Valcarce, A. A. R. 2021, *A&A*, 654, A138.

Molinaro, R., Ripepi, V., Marconi, M., Romaniello, M., Catanzaro, G., Cusano, F., De Somma, G., Musella, I., Storm, J., & Trentin, E. 2023, *MNRAS*, 520, 4154.

Mullen, J. P., Marengo, M., Martínez-Vázquez, C. E., Chaboyer, B., Bono, G., Braga, V. F., Dall'Ora, M., D'Orazi, V., Fabrizio, M., Monelli, M., & Thévenin, F. 2023, *ApJ*, 945, 83.

Muraveva, T., Delgado, H. E., Clementini, G., Sarro, L. M., & Garofalo, A. 2018, *MNRAS*, 481, 1195.

Nardetto, N., Mérand, A., Mourard, D., Storm, J., Gieren, W., Fouqué, P., Gallenne, A., Graczyk, D., Kervella, P., Neilson, H., Pietrzynski, G., Pilecki, B., Breitfelder, J., Berio, P., Challouf, M., Clausse, J. M., Ligi, R., Mathias, P., Meilland, A., Perraut, K., Poretti, E., Rainer, M., Spang, A., Stee, P., Tallon-Bosc, I., & ten Brummelaar, T. 2016, *A&A*, 593, A45.

Neeley, J. R., Marengo, M., Freedman, W. L., Madore, B. F., Beaton, R. L., Hatt, D., Hoyt, T., Monson, A. J., Rich, J. A., Sarajedini, A., Seibert, M., & Scowcroft, V. 2019, *MNRAS*, 490, 4254–4270.

Ngeow, C.-C., Bhardwaj, A., Dekany, R., Duev, D. A., Graham, M. J., Groom, S. L., Mahabal, A. A., Masci, F. J., Medford, M. S., & Riddle, R. 2022,a *AJ*, 163a, 239.

Ngeow, C.-C., Bhardwaj, A., Henderson, J.-Y., Graham, M. J., Laher, R. R., Medford, M. S., Purdum, J., & Rusholme, B. 2022,b *AJ*, 164b, 154.

O'Donnell, J. E. 1994, *ApJ*, 422, 158.

Owens, K. A., Freedman, W. L., Madore, B. F., & Lee, A. J. 2022, *ApJ*, 927, 8.

Pietrukowicz, P., Soszyński, I., & Udalski, A. 2021, *AcA*, 71, 205.

Pietrukowicz, P., Udalski, A., Soszyński, I., Skowron, D. M., Wrona, M., Szymański, M. K., Poleski, R., Ulaczyk, K., Kozłowski, S., Skowron, J., Mróz, P., Rybicki, K., Iwanek, P., & Gromadzki, M. 2020, *AcA*, 70, 121.

Ren, F., Chen, X., Zhang, H., de Grijs, R., Deng, L., & Huang, Y. 2021, *ApJL*, 911, L20.

Riess, A. G., Breuval, L., Yuan, W., Casertano, S., Macri, L. M., Bowers, J. B., Scolnic, D., Cantat-Gaudin, T., Anderson, R. I., & Cruz Reyes, M. 2022, *ApJ*, 938, 36.

Riess, A. G., Casertano, S., Anderson, J., MacKenty, J., & Filippenko, A. V. 2014, *ApJ*, 785, 161.

Riess, A. G., Casertano, S., Yuan, W., Bowers, J. B., Macri, L., Zinn, J. C., & Scolnic, D. 2021,a *ApJL*, 908a, L6.

Riess, A. G., Casertano, S., Yuan, W., Bowers, J. B., Macri, L., Zinn, J. C., & Scolnic, D. 2021,b *ApJL*, 908b, L6.

Riess, A. G., Casertano, S., Yuan, W., Macri, L., Anderson, J., MacKenty, J. W., Bowers, J. B., Clubb, K. I., Filippenko, A. V., Jones, D. O., & Tucker, B. E. 2018,a *ApJ*, 855a, 136.

Riess, A. G., Casertano, S., Yuan, W., Macri, L., Bucciarelli, B., Lattanzi, M. G., MacKenty, J. W., Bowers, J. B., Zheng, W., Filippenko, A. V., Huang, C., & Anderson, R. I. 2018,b *ApJ*, 861b, 126.

Ripepi, V., Catanzaro, G., Clementini, G., De Somma, G., Drimmel, R., Leccia, S., Marconi, M., Molinaro, R., Musella, I., & Poggio, E. 2022, *A&A*, 659, A167.

Ripepi, V., Clementini, G., Molinaro, R., Leccia, S., Plachy, E., Molnár, L., Rimoldini, L., Musella, I., Marconi, M., Garofalo, A., Audard, M., Holl, B., Evans, D. W., Jevardat de Fombelle, G., Lecoeur-Taibi, I., Marchal, O., Mowlavi, N., Muraveva, T., Nienartowicz, K., Sartoretti, P., Szabados, L., & Eyer, L. 2023, *A&A*, 674, A17.

Romaniello, M., Riess, A., Mancino, S., Anderson, R. I., Freudling, W., Kudritzki, R.-P., Macrì, L., Mucciarelli, A., & Yuan, W. 2022, *A&A*, 658, A29.

Skowron, D. M., Skowron, J., Mróz, P., Udalski, A., Pietrukowicz, P., Soszyński, I., Szymański, M. K., Poleski, R., Kozłowski, S., Ulaczyk, K., Rybicki, K., Iwanek, P., Wrona, M., & Gromadzki, M. 2019, *AcA*, 69, 305.

Skowron, D. M., Skowron, J., Udalski, A., Szymański, M. K., Soszyński, I., Wyrzykowski, Ł., Ulaczyk, K., Poleski, R., Kozłowski, S., Pietrukowicz, P., Mróz, P., Rybicki, K., Iwanek, P., Wrona, M., & Gromadzki, M. 2021, *ApJS*, 252, 23.

Stassun, K. G. & Torres, G. 2021, *ApJL*, 907, L33.

Trahin, B. 2019,. *Étalonnage de l'échelle des distances dans l'ère Gaia: Les étoiles pulsantes RR Lyrae et Céphéides*. PhD thesis, L'Université PSL, l'Observatoire de Paris.

Trahin, B., Breuval, L., Kervella, P., Mérand, A., Nardetto, N., Gallenne, A., Hocdé, V., & Gieren, W. 2021, *A&A*, 656, A102.

Trentin, E., Ripepi, V., Catanzaro, G., Storm, J., Marconi, M., De Somma, G., Testa, V., & Musella, I. 2023, *MNRAS*, 519, 2331.

van Leeuwen, F. 2007, *A&A*, 474, 653.

van Leeuwen, F. 2008, *VizieR Online Data Catalog*, 1311.

van Leeuwen, F., Feast, M. W., Whitelock, P. A., & Laney, C. D. 2007, *MNRAS*, 379, 723.

Vasiliev, E. & Baumgardt, H. 2021, *MNRAS*, 505, 5978–6002.

Wang, C., Yuan, H., & Huang, Y. 2022, *AJ*, 163, 149.

Wang, S. & Chen, X. 2023, *ApJ*, 946, 43.

Wang, S., Chen, X., de Grijs, R., & Deng, L. 2018, *ApJ*, 852, 78.

Wielgórski, P., Pietrzyński, G., Pilecki, B., Gieren, W., Zgirski, B., Górski, M., Hajdu, G., Narloch, W.,
    Karczmarek, P., Smolec, R., Kervella, P., Storm, J., Gallenne, A., Breuval, L., Lewis, M., Kałuszyński,
    M., Graczyk, D., Pych, W., Suchomska, K., Taormina, M., Rojas Garcia, G., Kotek, A., Chini, R.,
    Pozo Nũnez, F., Noroozi, S., Sobrino Figaredo, C., Haas, M., Hodapp, K., Mikołajczyk, P., Kotysz,
    K., Moździerski, D., & Kołaczek-Szymański, P. 2022, *ApJ*, 927, 89.

Zgirski, B., Pietrzyński, G., Górski, M., Gieren, W., Wielgórski, P., Karczmarek, P., Hajdu, G., Lewis, M.,
    Chini, R., Graczyk, D., Kałuszyński, M., Narloch, W., Pilecki, B., Rojas García, G., Suchomska, K.,
    & Taormina, M. 2023, *arXiv e-prints*,, arXiv:2305.09414.

Zhou, X. & Chen, X. 2021, *MNRAS*, 504, 4768–4784.

Zinn, J. C. 2021, *AJ*, 161, 214.

## Discussion

**Question (Anderson):** In the comparison of the cluster parallaxes you based your conclusions on Mauricio Cruz Reyes' membership list to determine the parallax of the cluster. What conclusion did you draw with respect to the zero-point offset relative to what he did? Are these things consistent or what do you see in there?

**Answer:** I didn't actually look at that in detail. I also, of course, applied the Lindegren correction, so I could make the direct comparison. I had no time to investigate that and I wanted to do the other comparison using no offset, just to make the message clear what happens. [The comparison of the average cluster parallaxes using other offsets is now included in these proceedings.]

**Question (Riess):** You're saying the countercorrections are worrisomely large but as it pertains to Cepheids, which you're looking at, the Cepheids are much brighter than where *Gaia* has done their calibration, so in that brightness range is it so surprising?

**Answer:** No, but if you have countercorrections of about $-20$ $\mu$as at 18 mag, it's almost like saying quasars are not at infinity. At the bright end, you see a large dispersion, as you know. We have to acknowledge that there is this 10 $\mu$as residual offset, no matter how you look at things. Countercorrections of $\pm 15$ $\mu$as, nobody can rule those out at the moment, to be fair.